\documentstyle[12pt,psfig]{article}
\makeatletter
\def\@cite#1#2{{[{#1}]\if@tempswa\typeout
{IJCGA warning: optional citation argument
ignored: `#2'} \fi}}

\newcount\@tempcntc
\def\@citex[#1]#2{\if@filesw\immediate\write\@auxout{\string\citation{#2}}\fi
  \@tempcnta\z@\@tempcntb\m@ne\def\@citea{}\@cite{\@for\@citeb:=#2\do
    {\@ifundefined
       {b@\@citeb}{\@citeo\@tempcntb\m@ne\@citea\def\@citea{,}{\bf ?}\@warning
       {Citation `\@citeb' on page \thepage \space undefined}}%
    {\setbox\z@\hbox{\global\@tempcntc0\csname b@\@citeb\endcsname\relax}%
     \ifnum\@tempcntc=\z@ \@citeo\@tempcntb\m@ne
       \@citea\def\@citea{,}\hbox{\csname b@\@citeb\endcsname}%
     \else
      \advance\@tempcntb\@ne
      \ifnum\@tempcntb=\@tempcntc
      \else\advance\@tempcntb\m@ne\@citeo
      \@tempcnta\@tempcntc\@tempcntb\@tempcntc\fi\fi}}\@citeo}{#1}}
\def\@citeo{\ifnum\@tempcnta>\@tempcntb\else\@citea\def\@citea{,}%
  \ifnum\@tempcnta=\@tempcntb\the\@tempcnta\else
   {\advance\@tempcnta\@ne\ifnum\@tempcnta=\@tempcntb \else \def\@citea{--}\fi
    \advance\@tempcnta\m@ne\the\@tempcnta\@citea\the\@tempcntb}\fi\fi}
\makeatother
\newenvironment{Eqnarray}%
     {\arraycolsep 0.14em\begin{eqnarray}}{\end{eqnarray}}

\def\simlt{\stackrel{<}{{}_\sim}}
\def\simgt{\stackrel{>}{{}_\sim}}
\def\be{\begin{equation}}
\def\ee{\end{equation}}
\def\bear{\be\begin{array}}
\def\eear{\end{array}\ee}
\def\bea{\begin{Eqnarray}}
\def\eea{\end{Eqnarray}}

\def\lsim{\mathrel{\raise.3ex\hbox{$<$\kern-.75em\lower1ex\hbox{$\sim$}}}}
\def\gsim{\mathrel{\raise.3ex\hbox{$>$\kern-.75em\lower1ex\hbox{$\sim$}}}}
\def\ifmath#1{\relax\ifmmode #1\else $#1$\fi}
\def\ls#1{\ifmath{_{\lower1.5pt\hbox{$\scriptstyle #1$}}}}

\def\beq{\begin{equation}}
\def\eeq{\end{equation}}
\def\beqa{\begin{Eqnarray}}
\def\eeqa{\end{Eqnarray}}

\def\baselinestretch{1}
\begin{document}
\def\IJMPA #1 #2 #3 {{\sl Int.~J.~Mod.~Phys.}~{\bf A#1}\ (19#2) #3$\,$}
\def\MPLA #1 #2 #3 {{\sl Mod.~Phys.~Lett.}~{\bf A#1}\ (19#2) #3$\,$}
\def\NPB #1 #2 #3 {{\sl Nucl.~Phys.}~{\bf B#1}\ (19#2) #3$\,$}
\def\PLB #1 #2 #3 {{\sl Phys.~Lett.}~{\bf B#1}\ (19#2) #3$\,$}
\def\PR #1 #2 #3 {{\sl Phys.~Rep.}~{\bf#1}\ (19#2) #3$\,$}
\def\JHEP #1 #2 #3 {{\sl JHEP}~{\bf #1}~(19#2)~#3$\,$}
\def\PRD #1 #2 #3 {{\sl Phys.~Rev.}~{\bf D#1}\ (19#2) #3$\,$}
\def\PTP #1 #2 #3 {{\sl Prog.~Theor.~Phys.}~{\bf #1}\ (19#2) #3$\,$}
\def\PRL #1 #2 #3 {{\sl Phys.~Rev.~Lett.}~{\bf#1}\ (19#2) #3$\,$}
\def\RMP #1 #2 #3 {{\sl Rev.~Mod.~Phys.}~{\bf#1}\ (19#2) #3$\,$}
\def\ZPC #1 #2 #3 {{\sl Z.~Phys.}~{\bf C#1}\ (19#2) #3$\,$}
\def\PPNP#1 #2 #3 {{\sl Prog. Part. Nucl. Phys. }{\bf #1} (#2) #3$\,$}

\catcode`@=11
\newtoks\@stequation
\def\subequations{\refstepcounter{equation}%
\edef\@savedequation{\the\c@equation}%
  \@stequation=\expandafter{\theequation}
  \edef\@savedtheequation{\the\@stequation}
  \edef\oldtheequation{\theequation}%
  \setcounter{equation}{0}%
  \def\theequation{\oldtheequation\alph{equation}}}
\def\endsubequations{\setcounter{equation}{\@savedequation}%
  \@stequation=\expandafter{\@savedtheequation}%
  \edef\theequation{\the\@stequation}\global\@ignoretrue

\noindent}
\catcode`@=12
\begin{titlepage}

\title{{\bf Oscillating neutrinos and $\mu\rightarrow e, \gamma$}}
\vskip2in \author{  {\bf J.A. Casas$^{1}$\footnote{\baselineskip=16pt
E-mail: {\tt alberto@makoki.iem.csic.es}}} and {\bf
A. Ibarra$^{1,2}$\footnote{\baselineskip=16pt  E-mail: {\tt
a.ibarra@physics.ox.ac.uk}}}\\
\hspace{3cm}\\
 $^{1}$~{\small Instituto de Estructura de la Materia, CSIC}\\ {\small
 Serrano 123, 28006 Madrid}
\hspace{0.3cm}\\ $^{2}$~{\small Department of Physics, Theoretical
Physics, University of Oxford}\\ {\small 1 Keble Road, Oxford OX1 3NP,
United Kingdom}.  } \date{} \maketitle \def\baselinestretch{1.15}
\begin{abstract}
\noindent

If neutrino masses and mixings are suitable to explain the atmospheric
and solar neutrino fluxes, this amounts to contributions to FCNC
processes, in particular  $\mu\rightarrow e, \gamma$. If the theory
is supersymmetric and the origin of the masses is a see-saw mechanism,
we show that the prediction for  BR($\mu\rightarrow e, \gamma$) is in
general larger than the experimental upper bound, especially if the
largest Yukawa coupling is ${\cal O}(1)$ and the
solar data are explained by a large angle MSW effect, which recent
analyses suggest as the preferred scenario. Our analysis is bottom-up
and completely general, i.e. it is based just on observable low-energy
data.  The work generalizes previous results
of the literature, identifying the dominant contributions. 
Application of the results to scenarios  with approximate
top-neutrino unification, like SO(10) models, rules out most of them
unless the leptonic Yukawa matrices satisfy very precise
requirements. Other possible ways-out, like gauge mediated SUSY
breaking, are also discussed.

\end{abstract}

\thispagestyle{empty} \leftline{} \leftline{}
\leftline{March 2001} \leftline{}

\vskip-22cm 
\rightline{} 
\rightline{IEM-FT-211/01}
\rightline{OUTP-01-11P}
\rightline{IFT-UAM/CSIC-01-08}
\rightline{hep-ph/0103065} \vskip3in

\end{titlepage}
\setcounter{footnote}{1} \setcounter{page}{1}
\newpage
\baselineskip=20pt

\noindent

\section{Introduction}

In the pure Standard Model, flavour is exactly conserved in the
leptonic sector since one can always choose a basis in which the
(charged) lepton Yukawa matrix, $\bf{Y_e}$, and gauge interactions are
flavour-diagonal. If neutrinos are massive and mixed, as suggested by
the observation of atmospheric and solar fluxes
\cite{SK}, this is no longer
true and there exists a source of lepton flavour violation (LFV), 
in analogy with the
Kobayashi--Maskawa mechanism in the quark sector.  Unfortunately, due
to the smallness of the neutrinos masses, the predicted branching
ratios for these processes are so tiny that they are completely
unobservable, namely  BR$(\mu \rightarrow e \gamma) < 10^{-50}$
\cite{meg_SM}.

In a supersymmetric (SUSY) framework the situation is completely different.
Besides the previous mechanism, supersymmetry provides new direct
sources of flavour violation in the leptonic sector, namely the
possible presence of off-diagonal soft terms
%
%
in the slepton mass matrices $\left(m_L^2\right)_{ij}$,
$\left(m_{e_R}^2\right)_{ij}$,  and trilinear couplings $A^e_{ij}$ 
\cite{offdiag}.
There are in the literature very strong bounds on these matrix
elements coming from requiring branching ratios for LFV processes to
be below the experimental rates \cite{fv}. The strongest bounds come precisely
from  BR($\mu \rightarrow e \gamma $).

The most drastic way to avoid these dangerous off-diagonal terms is to
impose perfect universality of the $\left(m_L^2\right)_{ij}$,
$\left(m_{e_R}^2\right)_{ij}$,  $A^e_{ij}$ matrices, i.e. to take them
proportional to the unit matrix.  Then, of course, they maintain this
form in any basis, in particular in the basis where all the gauge
interactions and $\bf{Y_e}$ are flavour-diagonal, so that we are left
just with the previous tiny Kobayashi--Maskawa LFV effects.  In
principle, there is no theoretical reason to impose the universality
constraint, and that is the so-called supersymmetric flavour
problem. However, there are some particular supersymmetric scenarios,
such as minimal supergravity, dilaton-dominated SUSY breaking or
gauge-mediated SUSY breaking, that give rise to the desired universal
soft terms (at least with much accuracy). In any case, the
universality assumption is obviously the most conservative supposition
that one can make about the form of the soft breaking terms when
analyzing flavour violation effects; and this will be our starting point.

It turns out, however, that even under this extremely conservative
assumption, if neutrinos are massive, radiative corrections may
generate off-diagonal soft terms. The reason is that the flavour
changing operators giving rise to the (non-diagonal) neutrino matrices
will normally contribute to the renormalization group equations (RGEs)
of the $\left(m_L^2\right)_{ij}$, $\left(m_{e_R}^2\right)_{ij}$ and
$A^e_{ij}$ matrices, inducing off-diagonal entries.

The most interesting example of this occurs when neutrino masses are
produced by a see-saw mechanism \cite{seesaw}. Then, above a certain scale $M$,
there are new degrees of freedom, the right-handed neutrinos, which
have conventional Yukawa couplings with a matricial Yukawa coupling
$\bf{Y_\nu}$. Normally, $\bf{Y_e}$ and $\bf{Y_\nu}$ cannot be
diagonalized simultaneously. Working in the flavour basis where
$\bf{Y_e}$ is diagonal, the  off-diagonal entries of $\bf{Y_\nu}$
(more precisely, the off-diagonal entries of  ${\bf Y_\nu^+}{\bf
Y_\nu}$) drive off-diagonal entries in the previous
$\left(m_L^2\right)_{ij}$, and $A^e_{ij}$ matrices through the RG
running \cite{Borzumati:1986qx}.

Actually, the supersymmetric see-saw is a extremely interesting
scenario  for a number of reasons. First, the see-saw mechanism is
probably the most convincing and economical mechanism to naturally
produce tiny neutrino masses (certainly it is the most popular
one). Second, a  {\em non}-supersymmetric see-saw suffers from a serious
hierarchy problem. Perhaps this point has not been sufficiently
acknowledged in the literature and we would like to emphasize it.  It
is common lore that the Standard Model (SM) presents a generic gauge
hierarchy problem. Namely, the Higgs mass acquires
quadratically-divergent radiative corrections, which, under the usual
interpretation of the renormalization process, are naturally of the
size of the scale at which new relevant physics enters. In a see-saw
scenario the problem becomes more acute. The presence of very massive
fermions (the right-handed neutrinos) coupled to the Higgs field
produces logarithmically-divergent radiative corrections to the Higgs
mass, which are proportional to $M^2$. This can be immediately noticed
from the usual expression of the 1-loop effective potential, where the
neutrino mass eigenstates which are essentially right-handed neutrinos
contribute as
$\frac{1}{64\pi^2}M^2Y_\nu^2H^2(\log\frac{M^2}{\mu^2}-\frac{3}{2})+
...$   ($\mu$ is the renormalization scale) \cite{vicente}
. In the supersymmetric
framework, this problem is automatically cured by the presence of the
right-handed sneutrinos with a similar mass. Consequently, it is fair
to say that supersymmetry is the most natural arena to implement the
see-saw mechanism.

This is precisely the scenario that we will consider in this paper,
computing BR($\mu \rightarrow e \gamma $) in the most general case
compatible with the observational (atmospheric and solar) indications
about neutrino masses and mixing angles. This subject has been
addressed in previous works, especially in refs.
\cite{topdown,Hisano:1996cp,Hisano:1999fj,bottomup}
(other related works can be found in ref.\cite{otros}). Let us
briefly comment on the differences between the treatment of these
works and the present one. The authors of ref.\cite{topdown} adopt a
``top-down'' point of view, i.e. they start with some sets of
particular textures for  $\bf{Y_e}$ and $\bf{Y_\nu}$, coming from more
or less well motivated symmetries, e.g. family symmetries,
 which can
produce sensible low-energy  neutrino mass matrices, $\cal{M_\nu}$, at
the  end of the day. Then, they study the corresponding predictions
for BR($\mu \rightarrow e \gamma $).  On the other hand, the authors
of ref.\cite{Hisano:1996cp,Hisano:1999fj,bottomup} adopt a 
``bottom-up'' point of view, i.e. they start
with the experimental data about $\cal{M_\nu}$ at low energy, finding
out a form of  $\bf{Y_\nu}$ able to reproduce these data, and then
computing the corresponding BR($\mu \rightarrow e \gamma $).  Both
approaches are interesting, but not general, as they are related to
particular choices of $\bf{Y_\nu}$. As a consequence, their results
are not conclusive about what can we expect for BR($\mu \rightarrow e
\gamma $) based on actual neutrino data.  In our case, we also follow
a ``bottom-up'' approach, i.e. we start with the existing experimental
information about $\cal{M_\nu}$.  Then we find the {\it most general}
textures for $\bf{Y_\nu}$ and $\cal{M}$ (the Majorana mass matrix of
the right-handed neutrinos) consistent with that information.  In our
opinion, this is by itself an interesting issue. The textures turn out
to depend on some free parameters, but very few (as we shall see, the
textures used in ref.\cite{Hisano:1999fj} correspond to a particular choice for
these parameters). Next, we compute BR($\mu \rightarrow e \gamma $)
and compare it with present and forthcoming experimental bounds. The
approach can be considered as a generalization of that by Hisano et al. in
ref.\cite{Hisano:1996cp,Hisano:1999fj}. As a matter of fact,
 we identify the dominant contributions, which were not considered 
in ref.\cite{Hisano:1996cp,Hisano:1999fj} as a
consequence of the particular texture used there.

The paper is organized as follows. In sect.~2 we determine the most
general form of  ${\bf Y_\nu}$ and ${\bf Y_\nu^+}{\bf Y_\nu}$,
compatible with all  the phenomenological requirements. In several
subsections we specialize the general formulas previously  obtained 
for relevant physical scenarios. In sect.~3 we review the way in which the
RG-running induces off-diagonal terms, and the relevance of the latter
for $l_i\rightarrow l_j, \gamma$ processes. In sect.~4 we apply the
information of sects. 2, 3 to compute the predictions on
BR($l_i\rightarrow l_j, \gamma$), with particular focus on
BR($\mu\rightarrow e, \gamma$). These predictions can be understood
analytically using some approximations, for the sake of clarity 
in the discussion. Again, in
several subsections we present and discuss the results for different
physical scenarios. In sect.~5 we offer a summary of the predictions
on BR($l_i\rightarrow l_j, \gamma$) obtained in the previous
sections. In sect.~6 we present our concluding remarks. Finally, we
include an Appendix with the most important formulas that we have used
for the full computations, specially RGEs and expressions for
BR($l_i\rightarrow l_j, \gamma$).

\section{General textures reproducing experimental data}

The supersymmetric version of the see-saw mechanism has a
superpotential
\bea
\label{superp}
W=W_{0} - \frac{1}{2}{\nu_R^c}^T{\cal M}\nu_R^c +{\nu_R^c}^T {\bf
Y_\nu} L\cdot H_2, \eea
where $W_{0}$ is the observable superpotential, except for neutrino masses, 
of the preferred version of the supersymmetric SM, e.g. the MSSM. The extra 
terms involve three additional neutrino chiral fields (one per generation;
indices are suppressed) not charged under the SM group: ${\nu_{R}}_i$
($i=e,\mu,\tau$). ${\bf Y_\nu}$ is the matrix of neutrino Yukawa
couplings,  $L_i$ ($i=e,\mu,\tau$) are the  left-handed lepton
doublets and $H_2$ is the hypercharge $+1/2$ Higgs doublet. The Dirac
mass matrix is given by ${\bf m_D}={\bf Y_\nu}\langle H_2^0\rangle$.
Finally, ${\cal M}$ is a $3\times 3$ Majorana mass matrix which does
not break the SM gauge symmetry. It is natural to assume that the
overall scale of ${\cal M}$, which we will denote  by $M$, is much
larger than the electroweak scale or any soft mass. Below $M$ the
theory is  governed by an effective superpotential
\bea W_{eff}=W_{0}+\frac{1}{2}({\bf Y_\nu}L\cdot H_2)^T{\cal
M}^{-1}({\bf Y_\nu}L\cdot H_2), \eea
obtained by integrating out the heavy neutrino fields in
(\ref{superp}).  The corresponding effective Lagrangian contains a
mass term for the left-handed neutrinos:
\bea  \delta {\cal L}=-\frac{1}{2}\nu^T{\cal M}_\nu \nu + {\rm h.c.},
\eea
with
\bea
\label{seesaw}
{\cal M}_\nu= {\bf m_D}^T {\cal M}^{-1} {\bf m_D} =  {\bf Y_\nu}^T
{\cal M}^{-1} {\bf Y_\nu} \langle H_2^0\rangle^2,   \eea
suppressed with respect to the typical fermion masses by the  inverse
power of the large scale $M$.  It is convenient to extract the Higgs
VEV by defining the $\kappa$ matrix as
\bea
\label{kappa}
\kappa = {\cal M}_\nu/ \langle H_2^0\rangle^2= {\bf Y_\nu}^T {\cal
M}^{-1} {\bf Y_\nu},  \eea
where $\langle H_2^0\rangle^2=v_2^2=v^2 \sin^2\beta$ and   $v=174$
GeV. The experimental data about neutrino masses and mixings are
referred to the ${\cal M}_\nu$ matrix, or equivalently $\kappa$,
evaluated at low energy (electroweak scale). In this sense, it should
be noted that eqs.(\ref{seesaw}, \ref{kappa}) are not defined at low
energy but at  the ``Majorana scale'', $M$. Therefore, in order to
compare to the experiment one has still to run $\kappa$ down to low
energy through the corresponding RGE\footnote{ For hierarchical
neutrinos, with which we will be mainly concerned, this running does
not substantially affect the texture of $\kappa$, and thus the mixing
angles. It does modify, however, the overall size of $\kappa$, which
has to be taken into account.}. Alternatively, following the bottom-up
spirit, one can start with the physical $\kappa$ matrix at low energy
and then run it upwards to the $M$ scale, where eqs.(\ref{seesaw},
\ref{kappa}) are defined. This has been the procedure we have followed
in all the numerical calculations to be presented in the next sections.

Working in the flavour basis in which the charged-lepton Yukawa
matrix, $\bf{Y_e}$, and gauge interactions are flavour-diagonal,  the
$\kappa$ matrix is diagonalized by the MNS \cite{MNS} matrix $U$ according to
\be
\label{Udiag}
U^T{\kappa } U={\mathrm diag}(\kappa_1,\kappa_2,\kappa_3)\equiv
D_\kappa, \ee
where $U$ is a unitary matrix that relates  flavour to mass eigenstates
\bea  \pmatrix{\nu_e \cr \nu_\mu\cr \nu_\tau\cr}= U \pmatrix{\nu_1\cr
\nu_2\cr \nu_3\cr}\,.
\label{CKM}
\eea
It is possible, and sometimes convenient, to choose  $\kappa_i\geq
0$. Then, $U$ can be written as
\bea U=V\cdot {\mathrm diag}(e^{-i\phi/2},e^{-i\phi'/2},1)\ \ ,
\label{UV}
\eea
where $\phi$ and $\phi'$ are CP violating phases (if different from
$0$ or $\pi$) and $V$ has the ordinary form of a CKM matrix
\be \label{Vdef} V=\pmatrix{c_{13}c_{12} & c_{13}s_{12} & s_{13}e^{-i\delta}\cr
-c_{23}s_{12}-s_{23}s_{13}c_{12}e^{i\delta} & c_{23}c_{12}-s_{23}s_{13}s_{12}e^{i\delta} & s_{23}c_{13}\cr
s_{23}s_{12}-c_{23}s_{13}c_{12}e^{i\delta} & -s_{23}c_{12}-c_{23}s_{13}s_{12}e^{i\delta} &
c_{23}c_{13}\cr}.  \ee
On the other hand, one can always choose to work in a basis of right
neutrinos where ${\cal M}$ is diagonal
\be {\cal M}={\mathrm diag}({\cal M}_1,{\cal M}_2,{\cal M}_3)\equiv
D_{\cal M}, 
\ee
with ${\cal M}_i\geq 0$. Then, from eqs.(\ref{kappa}, \ref{Udiag})
\bea
\label{previa}
D_\kappa= U^T{\bf Y_\nu}^TD_{{\cal M}^{-1}}{\bf Y_\nu}U=  U^T{\bf
Y_\nu}^TD_{\sqrt{{\cal M}^{-1}}} D_{\sqrt{{\cal M}^{-1}}} {\bf Y_\nu}U
\eea
where, in an obvious notation, $D_{\sqrt{A}}\equiv +\sqrt{D_{A}}$.
Multiplying both members of the eq.(\ref{previa})  
by  $D_{\sqrt{\kappa^{-1}}}$
from the left and from the
right, we get
\bea
{\bf 1}= \left[  D_{ \sqrt{{\cal M}^{-1}}} {\bf Y_\nu}U
D_{\sqrt{\kappa^{-1}}} \right]^T \left[  D_{ \sqrt{{\cal M}^{-1}}}
{\bf Y_\nu}U D_{\sqrt{\kappa^{-1}}} \right] \eea
whose solution is $D_{\sqrt{{\cal M}^{-1}}} {\bf Y_\nu}U
D_{\sqrt{\kappa^{-1}}}=R$, with $R$ any orthogonal matrix ($R$ can be
complex provided $R^TR= {\bf 1}$).  Hence, in order to reproduce the
physical, low-energy, parameters, i.e. the light neutrino masses
(contained in $D_{\kappa}$) and mixing angles and CP phases (contained
in $U$), the most general ${\bf Y_\nu}$ matrix is given by
\bea
\label{Ynu}
{\bf Y_\nu}=D_{\cal \sqrt{M}} R D_{\sqrt{\kappa}} U^+  
\eea
So, besides the physical and measurable low-energy parameters
contained in $D_{\kappa}$ and $U$, ${\bf Y_\nu}$ depends on the three
(unknown)  positive mass eigenvalues of the righthanded neutrinos and
on the three  (unknown) complex parameters defining
$R$. We will see, however, that in practical cases the number  of
relevant free parameters becomes drastically reduced.  Let us stress
that eq.(\ref{Ynu}), which is central for us, has to be understood at
the $M$ scale and in the above-defined basis, i.e.  the flavour basis
in which the charged-lepton Yukawa matrix, $\bf{Y_e}$, the
right-handed Majorana  mass matrix, ${\cal M}$, and the gauge
interactions are flavour-diagonal.  The extension of eq.(\ref{Ynu}) to
other choices of basis is straightforward,  taking into account the
transformation properties of the $\bf{Y_e}$,  ${\bf Y_\nu}$, ${\cal
M}$ matrices under changes of basis. For example,  if one works in a
basis of ${\nu_R}$ where the  ${\cal M}$ matrix is  non-diagonal, then
the most general ${\bf Y_\nu}$ matrix reads  ${\bf Y_\nu}=U_M^*D_{\cal
\sqrt{M}} R D_{\sqrt{\kappa}} U^+$, where $U_M$ is such that ${\cal
M}=U_M^* D_{\cal M} U_M^+$.  Let us also mention that eq.(\ref{Ynu})
is valid as well for the non-supersymmetric see-saw mechanism.

As commented in sect.~1, in ref.\cite{Hisano:1999fj}
 a particular choice for  ${\bf
Y_\nu}$ was taken, which corresponds to take $R={\bf 1}$ in
eq.(\ref{Ynu}).  This is equivalent to assume that there exists a basis
of $L_i$  and ${\nu_R}_i$ in which ${\bf Y_\nu}$ and ${\cal M}$ are
simultaneously diagonal (though not ${\bf Y_e}$). To see this notice
that if $R={\bf 1}$, then ${\bf Y_\nu}$, as given by eq.(\ref{Ynu}),
can be made diagonal by rotating $L_i$ with the $U$-matrix. This
hypothesis may be  consistent with certain models,  namely when all
the leptonic flavour violation can be attributed to the sector of
charged leptons, but clearly it is not  general, and it is
inconsistent with other scenarios.

Another special choice of $R$ occurs when the ${\bf Y_\nu}$ matrix
given by eq.(\ref{Ynu}) has the form ${\bf Y_\nu}= W D_Y$, where $D_Y$
is a diagonal matrix and $W$ is a unitary matrix. From the see-saw
formula (\ref{kappa}), $W$ is related to $D_{\cal M}$ and $D_Y$
through $D_{ \sqrt{{\cal M}^{-1}}} = W^* (D_{ Y^{-1}}\kappa D_{
Y^{-1}}) W^+$.  Then, by rotating  ${\nu_R}_i$ with the $W$-matrix,
$\bf{Y_e}$ and ${\bf Y_\nu}$ get  simultaneously diagonal, while
${\cal M}$ gets non-diagonal, more  precisely ${\cal
M}=D_Y\kappa^{-1}D_Y$. In other  words, all the leptonic flavour
violation can be attributed to the sector of right-handed neutrinos,
in contrast to the previous situation.
It is important to remark that both are very special situations, 
not generic at all.

On the other hand, as we will see in the next section, the rate  for
$l_i\rightarrow l_j, \gamma$ processes just depends on the  ${\bf
Y_\nu^+}{\bf Y_\nu}$ matrix (to be precise, on the matrix element
$({\bf Y_\nu^+}{\bf Y_\nu})_{ij}$) rather than on ${\bf Y_\nu}$,
since this is the quantity that
enters the RGEs of the off-diagonal soft parameters. From
eq.(\ref{Ynu}) ${\bf Y_\nu^+}{\bf Y_\nu}$ has the general form
\bea
\label{Ynu+Ynu}
{\bf Y_\nu^+}{\bf Y_\nu}= U D_{\sqrt{\kappa}} R^+ D_{\cal M} R
D_{\sqrt{\kappa}} U^+   \eea
Notice that ${\bf Y_\nu^+}{\bf Y_\nu}$, and therefore
eq.(\ref{Ynu+Ynu}), do not depend on the ${\nu_R}$-basis, and thus on
the fact that ${\cal M}$ is diagonal or not.

In order to illustrate the use of eqs.(\ref{Ynu}, \ref{Ynu+Ynu}),  let
us consider some interesting scenarios that often appear in the
literature

\subsection{$\nu_L$'s and $\nu_R$'s completely hierarchical}

In this case $\kappa_1\ll\kappa_2\ll\kappa_3$ and  ${\cal M}_1\ll{\cal
M}_2\ll{\cal M}_3$, so that $D_\kappa$ and $D_{\cal M}$ can be
approximated by
\be
\label{jerar}
D_\kappa\simeq {\mathrm diag}(0,\kappa_2,\kappa_3),\;\;\;\;  D_{\cal
{M}}\simeq {\mathrm diag}(0,0,{\cal M}_3) \ee
where $(v_2^2\kappa_2)^2\simeq (\Delta m_\nu^2)_{sol}$,
$(v_2^2\kappa_3)^2\simeq (\Delta m_\nu^2)_{atm}$. Then, from
eq.(\ref{Ynu}),
\bea
\label{Ynujer}
({\bf Y_\nu})_{ij} \simeq  \sqrt{ {\cal M}_3} \delta_{i3} R_{3l}
\sqrt{\kappa_l} U^+_{lj} 
\eea
So, for a certain set of low-energy physical quantities (given by
$\kappa_l$ and $U$), ${\bf Y_\nu}$ just depends on two (generally
complex)  parameters, $\sqrt{{\cal M}_3}R_{32}$ and $\sqrt{{\cal
M}_3}R_{33}$. Let us remark that, although the $({\bf Y_\nu})_{1j}$ and
$({\bf Y_\nu})_{2j}$ entries are proportional to  $\sqrt{{\cal M}_1}$ and
$\sqrt{{\cal M}_2}$ respectively, and thus suppressed, they are 
normally relevant
if one wishes to reconstruct the neutrino mass matrix, $\kappa$, from 
${\bf Y_\nu}$, $D_{\cal M}$ through the see-saw equation (\ref{kappa}).
Still, they are {\em not} relevant for the ${\bf Y_\nu^+}{\bf Y_\nu}$ 
entries, and thus for the predictions on BR($l_i\rightarrow
l_j, \gamma$), which is our main goal.

It is interesting to notice that this scenario also includes the
possibility  of an ``inverse hierarchy'', i.e. $\sim
{\mathrm diag}({\cal M}_3,0,0)$, or any other ordering.  The reason is
that one can permute the entries of $D_{\cal {M}}$ by  redefining the
$\nu_R$'s  through an appropriate rotation. In particular, one can
pass from an inverse hierarchy to an ordinary one through a rotation
in the 1--3 ``plane'': $\nu_R = R_p \nu_R'$ with $R_p=\left\{(0,0,1),\
(0,1,0),\ (-1,0,0)  \right\}$. Then, ${\bf Y_\nu}\rightarrow
{\bf Y_\nu}'=D_{\cal \sqrt{M}}'R_p^T R D_{\sqrt{\kappa}} U^+$
with $D_{\cal {M}}'= R_p^T D_{\cal {M}} R_p$. Hence, the general 
formula (\ref{Ynu})
and its particularization (\ref{Ynujer}) remain identical with the
modification $R \rightarrow R'=R_p^T  R$. Since $R$ runs over all
possible orthogonal matrices, so $R'$ does. So, both starting points
(inverse or ordinary hierarchy) lead to the same complete results.

In many models one can impose an extra condition coming from the
unification of Yukawa couplings. In particular, in simple $SO(10)$
models, usually one of the  ${\bf Y_\nu^+}{\bf Y_\nu}$ eigenvalues
(the largest one) coincides with the top Yukawa coupling $|Y_t|^2$ at
the unification scale, $M_X$ (in other more elaborate models, it
coincides with $3|Y_t|^2$) \cite{Mohapatra:1999vv}. 
We will denote $|Y_0|^2$ this
maximum eigenvalue.  Now, from eq.(\ref{Ynujer}), the largest
eigenvalue of ${\bf Y_\nu^+}{\bf Y_\nu}$ is $|Y_0|^2\simeq{\cal
M}_3\left( |R_{32}|^2 \kappa_2 + |R_{33}|^2 \kappa_3 \right)$,  which
(after running up to $M_X$) would be identified with $|Y_t(M_X)|^2$.
This eliminates one real parameter, leaving just $R_{33}/R_{32}$, plus
one phase, e.g. ${\mathrm phase}(R_{33})$, as the independent
parameters. Actually, as mentioned above, the $l_i\rightarrow l_j,
\gamma$ processes just depends on the  ${\bf Y_\nu^+}{\bf Y_\nu}$
matrix, which does not depend on the previous phase. Therefore, one is
just left with one (complex) parameter, $R_{33}/R_{32}$.

A convenient way to visualize this is to take the following 
parametrization\footnote{The parametrization (\ref{R}) does not 
count all the possible 
forms of $R$, as it does not include ``reflections''. However, 
for the scenario considered in this subsection it is general enough, 
since the uncounted forms of $R$ do not produce different textures
of ${\bf Y_\nu}$.} of $R$
%
%
\bea
\label{R}
R=\pmatrix{\hat c_2\hat c_3 & -\hat c_1\hat s_3-\hat s_1\hat s_2\hat
c_3  & \hat s_1\hat s_3-\hat c_1\hat s_2\hat c_3 \cr  \hat c_2\hat s_3
& \hat c_1\hat c_3-\hat s_1\hat s_2\hat s_3   & -\hat s_1\hat c_3-\hat
c_1\hat s_2\hat s_3 \cr \hat s_2  & \hat s_1\hat c_2 & \hat c_1\hat
c_2\cr}\;,  \eea
where $\hat \theta_1$, $\hat \theta_2$, $\hat \theta_3$ are arbitrary
complex angles. Then, from eq.(\ref{Ynujer})
\bea
\label{Ynujer2}
{\bf Y_\nu}&\simeq&\frac{|Y_0|}{\sqrt{( |\hat s_1|^2 \kappa_2 + |\hat
c_1|^2  \kappa_3}} \frac{\hat c_2}{|\hat c_2|}\nonumber \\ &\times&
\pmatrix{\hat 0& 0& 0\cr 0&0&0\cr \hat s_1 \sqrt{\kappa_2}U^*_{12} +
\hat c_1 \sqrt{\kappa_3}U^*_{13} & \hat s_1 \sqrt{\kappa_2}U^*_{22} +
\hat c_1 \sqrt{\kappa_3}U^*_{23} &\hat s_1 \sqrt{\kappa_2}U^*_{32} +
\hat c_1 \sqrt{\kappa_3}U^*_{33}}   \eea
which depends just on the angle $\hat \theta_1$ and the phase
$\frac{\hat c_2}{|\hat c_2|}$; ${\bf Y_\nu^+}{\bf Y_\nu}$ depends just
on $\hat \theta_1$. The previous equation should be understood at the
${\cal M}$ scale; the prefactor $|Y_0|$ represents the square root  of
the maximal eigenvalue of ${\bf Y_\nu^+}{\bf Y_\nu}$ at  the ${\cal
M}$-scale.

In the typical case in which everything is real, very used in the
literature, the $\hat \theta_i$ angles are real, and hence ${\bf
Y_\nu}$ and ${\bf Y_\nu^+}{\bf Y_\nu}$ depend on a single real 
parameter,  $\hat
\theta_1$. Therefore, in spite of the additional degrees of freedom
introduced by the lack of
knowledge about $R$, this scenario is quite
predictive.

In the previous discussion, it has been assumed that at least one of
the  two entries, $R_{32}$ or $R_{33}$, is different from zero.
However, it may happen that
$R_{32}=R_{33}=0$. Then, $\kappa_1,\ {\cal M}_1,\ {\cal M}_2$  cannot
be neglected anymore in the general equation eq.(\ref{Ynu}).   It is
worth-noticing that going to the basis where the $D_{\cal {M}}$ matrix
is `inverse hierarchical', i.e.  $\nu_R \rightarrow R_p \nu_R$,
$D_{\cal {M}}\rightarrow {\mathrm diag}({\cal M}_3,{\cal M}_2,{\cal
M}_1)$, this possibility  translates into $R_{12}=R_{13}=0$,  which
includes the particularly simple case $R={\bf 1}$.

\subsection{$\nu_L$'s hierarchical and $\nu_R$'s degenerate}

In this case,  $D_\kappa$ and $D_{\cal M}$ have the form
\be
\label{degen}
D_\kappa\simeq {\mathrm diag}(0,\kappa_2,\kappa_3),\;\;\;\;  D_{\cal
{M}}\simeq {\mathrm diag}({\cal M},{\cal M},{\cal M})   \ee
where, again, $(v_2^2\kappa_2)^2\simeq (\Delta m^2)_{sol}$,
$(v_2^2\kappa_3)^2\simeq (\Delta m^2)_{atm}$. Then, from
eq.(\ref{Ynu}),
\bea
\label{Ynudeg}
({\bf Y_\nu})_{ij} \simeq  \sqrt{ {\cal M}} R_{il} \sqrt{\kappa_l}
U^+_{lj}  \eea
So, in principle, for a certain set of low-energy physical quantities
(given by $\kappa_q$ and $U$), ${\bf Y_\nu}$ depends on one real
parameter (${\cal M}$) and three complex ones 
defining the $R$ matrix. Notice,
however,  that in the real case, where $R^+=R^T$, ${\bf Y_\nu^+}{\bf
Y_\nu}$ takes  the simple form
\bea
\label{Ynu+Ynu2}
{\bf Y_\nu^+}{\bf Y_\nu}={\cal M} U D_{\kappa} U^+ \ , \eea
which just contains one free parameter, ${\cal M}$. This can be fixed
by a unification condition, similarly to the previous case. Here
$|Y_0|^2 \simeq {\cal M}\kappa_3\simeq |Y_t|^2$ (where the
identification  must be understood at the  $M_X$ scale). Therefore,
this is also a very predictive scenario.  Notice that here  ${\bf
Y_\nu^+}{\bf Y_\nu}$ does not depend on the form of $R$, so in this
case the choice $R={\bf 1}$ is indeed general enough to study 
$l_i\rightarrow
l_j, \gamma$ processes.

\subsection{$\nu_L$'s quasi-degenerate}

In this case $D_\kappa\simeq {\mathrm
diag}(\kappa_1,\kappa_2,\kappa_3)$,  with
$\kappa_1\sim\kappa_2\sim\kappa_3\equiv \kappa$.  Then, it is logical
to assume that ${\cal M}$ has degenerate eigenvalues, otherwise a big
conspiracy would be  needed between ${\bf Y_\nu}$ and ${\cal M}$ so
that $\kappa_1\sim\kappa_2\sim\kappa_3$ in eq.(\ref{kappa}). Hence
\be
\label{degenL}
D_\kappa\simeq {\mathrm diag}(\kappa,\kappa,\kappa),\;\;\;\;  D_{\cal
{M}}\simeq {\mathrm diag}({\cal M},{\cal M},{\cal M})\;.   \ee
Consequently, eq.(\ref{Ynudeg}) and eq.(\ref{Ynu+Ynu2}) (for $R$ real) 
hold in
this case too.  Since $D_\kappa\simeq \kappa {\bf 1}$, from
eq.(\ref{Ynu+Ynu2}) we notice
\bea
\label{Ynu+Ynu22}
{\bf Y_\nu^+}{\bf Y_\nu}={\cal M} U D_{\kappa} U^+  \simeq |Y_0|^2
{\bf 1}\ , \eea
where $|Y_0|^2={\cal M} \kappa_3$ is the largest eigenvalue of the
${\bf Y_\nu^+}{\bf Y_\nu}$ matrix. This means that for
quasi-degenerate neutrinos, with $R$ real, we expect small
off-diagonal entries in ${\bf Y_\nu^+}{\bf Y_\nu}$ and thus suppressed
$l_i\rightarrow l_j, \gamma$ rates.  We will be more quantitative
about the size of those entries in subsect.~4.3.

The scenario of partially degenerate neutrinos will also be addressed
in that section.

\section{RG-induced lepton flavour violating soft terms and 
$l_i\rightarrow l_j, \gamma$}

In addition to the supersymmetric Lagrangian, which contains the gauge
interactions plus the part derived from the superpotential
(\ref{superp}), one has to consider the soft breaking terms (gaugino
and scalar masses, and trilinear and bilinear scalar terms) coming
from the (unknown) supersymmetry breaking mechanism.  In a
self-explanatory notation, they have the form
\bea  -{\cal L}_{\rm soft}=\ \left(m_L^2\right)_{ij} \bar L_iL_j\ +\
\left(m_{e_R}^2\right)_{ij} {\bar e}_{Ri} e_{Rj}\ + \ \left({A_e}_{ij}
{e_R^c}_i H_1 L_j +{\rm h.c.}\right)\  +\ {\rm etc.}\; ,  \eea
where we have written explicitly just the soft breaking terms in the
leptonic sector, namely scalar masses and trilinear scalar terms. All
the fields in the previous equation denote just the corresponding
scalar components. As explained in sect.~1, concerning flavour
violation the most conservative starting point for ${\cal L}_{\rm
soft}$ is the  assumption of universality, which corresponds to take
\bea
\label{universal} 
\left(m_L^2\right)_{ij} = m_0^2\ {\bf 1},\;\;\;
\left(m_{e_R}^2\right)_{ij} = m_0^2\ {\bf 1},\;\;\; {A_e}_{ij}=A_0\
{\bf Y_e}_{ij}\;,  \eea
so that working in the $L_i$ and ${e_R}_i$ basis where  ${\bf Y_e}$ is
diagonal, the soft terms do not contain off-diagonal (lepton flavour
violating) entries.

As mentioned in sect.~1, this is only strictly true at the scale where
universality is imposed, e.g. $M_X$ in GUT models. Below that scale,
the RGEs of the soft terms, which contain non-diagonal contributions
proportional to  ${\bf Y_\nu^+}{\bf Y_\nu}$, induce off-diagonal soft
terms\footnote{We will pay attention just to the contribution that
appears in the MSSM, i.e. the one $\propto {\bf Y_\nu^+}{\bf
Y_\nu}$. This is a minimal possibility. In more complicated models,
there can be additional sources of flavour violation in the RGEs,
leading to larger predictions for BR($l_i\rightarrow l_j, \gamma$).}.
These contributions are decoupled at the characteristic scale
of the right-handed neutrinos, $M$.  As noted e.g. 
in ref. \cite{Borzumati:1986qx},
and as it is clear from eqs.(\ref{rgmL}--\ref{rgAe}) 
in the Appendix, this mechanism is
not efficient to generate off-diagonal entries in the
$\left(m_{e_R}^2\right)$ matrix, since the corresponding RGE does  not
contain $\sim {\bf Y_\nu^+}{\bf Y_\nu}$ terms.  More precisely,  in
the leading-log approximation\footnote{We use the leading-log
approximation through the text in order to make the results easily
understandable. Nevertheless, the numerical results, to be exposed below,
have been obtained by integrating the full set of RGEs.}, the
off-diagonal  soft terms at low-energy are given by
\bea
\label{softafterRG} 
\left(m_L^2\right)_{ij} & \simeq & \frac{-1}{8\pi^2}(3m_0^2 + A_0^2)
({\bf Y_\nu^+}{\bf Y_\nu})_{ij} \log\frac{M_X}{M}\ , \nonumber\\
\left(m_{e_R}^2\right)_{ij} & \simeq & 0\ , \nonumber\\  ({A_e})_{ij}&
\simeq &   \frac{-3}{8\pi^2} A_0 Y_{l_i} ({\bf Y_\nu^+}{\bf
Y_\nu})_{ij} \log\frac{M_X}{M}\;,   \eea
where $i\neq j$ and $Y_{l_i}$ is the Yukawa coupling of the  charged
lepton $l_i$.

The amplitude for a $l_i\rightarrow l_j, \gamma$ process has the
general form
\bea
\label{T}
T=\epsilon^\alpha \bar{l_j} {m_l}_i i \sigma_{\alpha \beta} q^\beta
\left(A_L P_L + A_R P_R \right)l_i\;, \eea
where $q$ is the momentum of the photon, $ P_{RL}=(1\pm \gamma_5)/2$
and $A_L$ ($A_R$) is the coefficient of the amplitude when the
incoming $l_i$ lepton is left (right), and thus the $l_j$ lepton is
right (left). The expressions for  $A_L$, $A_R$ can be found in the
literature \cite{Hisano:1996cp} (see Appendix for more details).   The
corresponding branching ratio is given by
\bea
\label{BR}
{\mathrm{BR}}(l_i\rightarrow l_j, \gamma) =
\frac{12\pi^2}{G_F^2} \left(|A_L|^2 + |A_R|^2 \right)   \eea
Notice that the $l_i\rightarrow l_j, \gamma$ process violates the
chiral  symmetries ${l_R}_k\rightarrow e^{i\alpha_k}{l_R}_k$. Under
the assumption of initial universality, this symmetry is violated just
by the Yukawa coupling of the corresponding lepton. Thus, $A_L$
($A_R$) is proportional to the mass of the righthanded lepton
involved, i.e. ${l_j}_R$ (${l_i}_R$). Since ${m_l}_j^2 \ll{m_l}_i^2$,
then $|A_L|^2 \ll |A_R|^2 $. This can also be checked by inspection of
the diagrams that contribute to $l_i\rightarrow l_j, \gamma$, taking
into account eq.(\ref{softafterRG}). It is interesting to note that 
the angular distribution of $l_j$ with respect to the polarization 
direction of  the incoming lepton ($P_{l_i}$) depends on the 
relative magnitudes of the coefficients $A_L$, $A_R$. 
The relation  $|A_L|^2 \ll |A_R|^2 $ 
predicted by the see-saw mechanism, and which does not hold in general
in other models of generation of neutrino masses, gives a characteristic 
($1-P_{l_i}\cos\theta$) distribution that could be measured by future 
experiments \cite{kuno}.

We have taken into account all contributions to BR($l_i\rightarrow
l_j, \gamma$), using the general expressions given in the literature,
in particular from  ref.~\cite{Hisano:1996cp}, as explained 
in the  Appendix. These
have not been obtained by using the mass-insertion approximation, but
by diagonalizing all the mass matrices involved in the task, i.e.  those
of (left and right) sleptons, charginos and neutralinos. The diagrams
have the form shown in Fig.~1. The precise form of  BR($l_i\rightarrow
l_j, \gamma$) that we have used in our computations is a rather
cumbersome expression, given in the Appendix.  However, for the sake
of the physical discussion it is interesting to think in the 
mass-insertion approximation to identify the dominant contributions. As
discussed in ref. \cite{Hisano:1999fj}, these correspond to the mass-insertion
diagrams enhanced by $\tan\beta$ factors. All of them are proportional
to ${m_L^2}_{ij}$, and have the generic form shown in Fig.~2.  So, in
all cases
\begin{figure}
\centerline{\hbox{
\psfig{figure=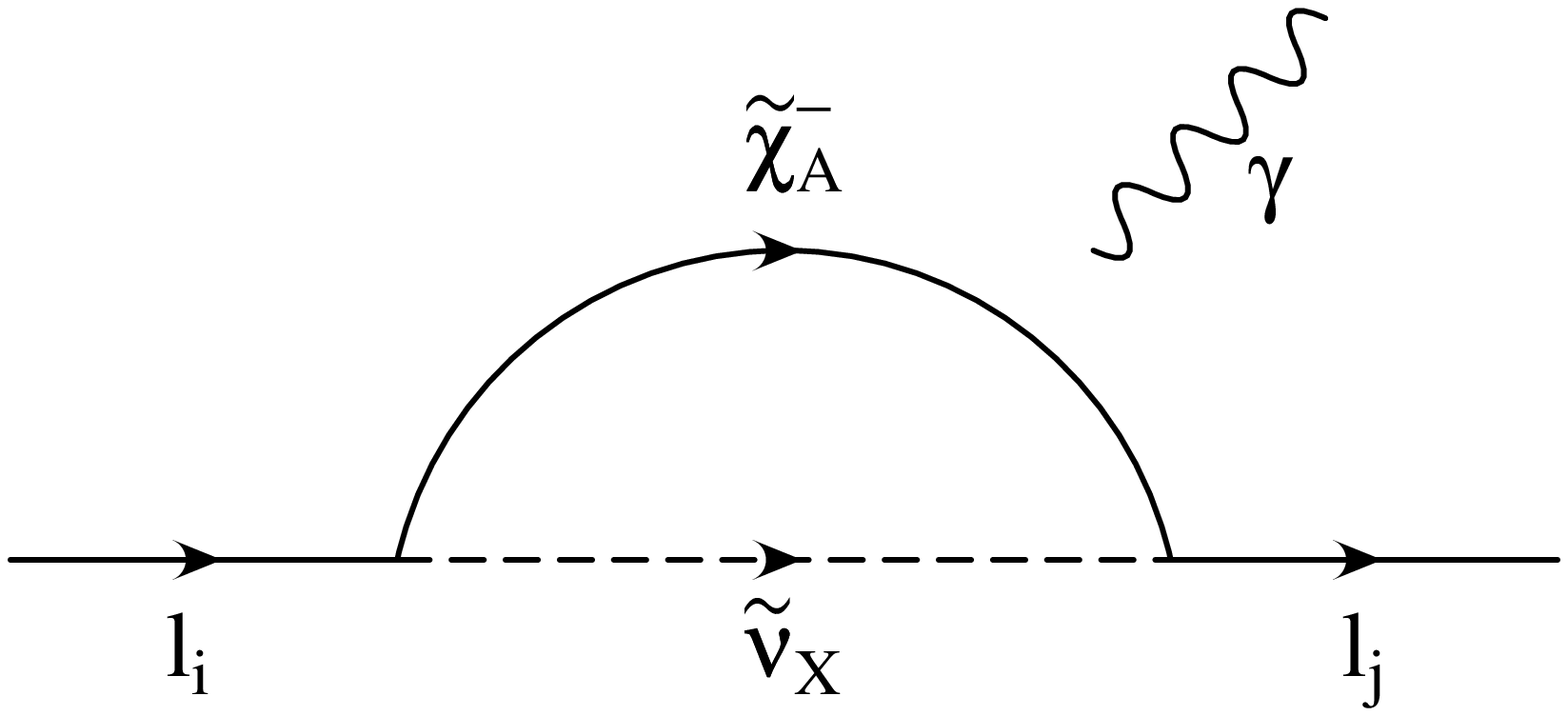,width=7.5cm,rwidth=8cm}
\psfig{figure=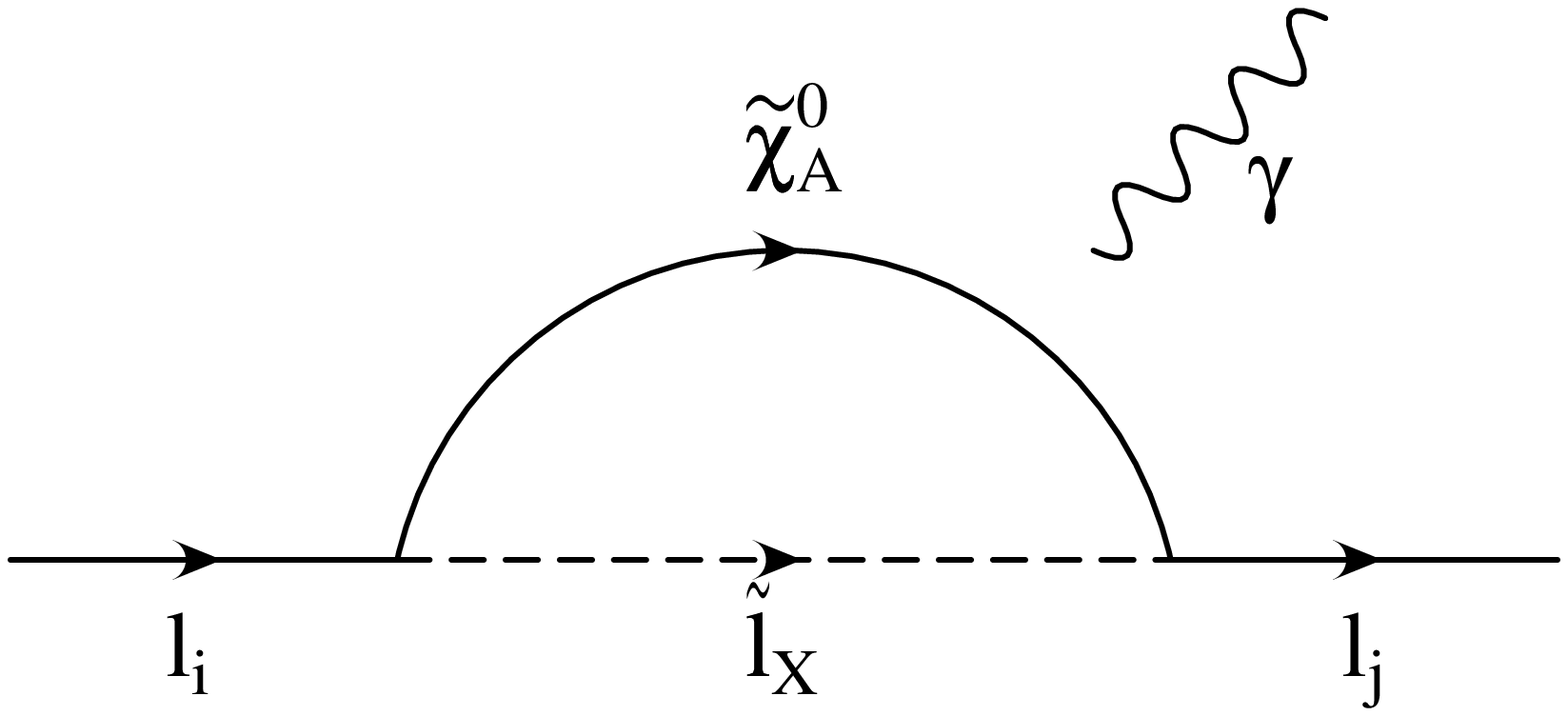,width=7.5cm}}}
\caption
{\footnotesize  Feynman diagrams contributing to 
$l_i\rightarrow l_j, \gamma$. $\tilde \nu_X$ and $\tilde l_X$
with $X=1,\cdots, 3$ ($4,\cdots, 6$) represent
the mass eigenstates of ``left'' (``right'') sneutrinos and charged sleptons, 
respectively.
$\tilde \chi^-_A$, $A=1, 2$, denote the charginos, whereas
$\tilde \chi^0_A$, $A=1,\cdots, 4$, denote the neutralinos.
}
\end{figure}

\begin{figure}
\centerline{\hbox{
\psfig{figure=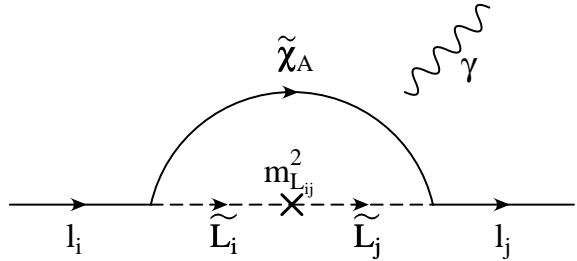,width=7.5cm}}}
\caption
{\footnotesize   Dominant 
Feynman diagrams contributing to 
$l_i\rightarrow l_j, \gamma$ in the mass-insertion approximation.
$\tilde L_i$ are the slepton doublets in the basis where 
the gauge interactions and the charged-lepton Yukawa couplings are 
flavour-diagonal. $\tilde \chi_A$ denote the charginos and neutralinos, 
as in Fig.~1.
}
\end{figure}

%
\bea
\label{BR2}
{\mathrm{BR}}(l_i\rightarrow l_j, \gamma) \simeq
\frac{12\pi^2}{G_F^2}|A_R|^2 \sim \frac{\alpha^3}{G_F^2}
\frac{{|m_L^2}_{ij}|^2}{m_S^8}\tan^2\beta \;,  \eea
where $m_S^2$ represents supersymmetric leptonic scalar masses.

From eq.(\ref{BR2}) it is clear that BR($l_i\rightarrow l_j, \gamma$)
depends crucially on the quantity ${m_L^2}_{ij}$, which at low energy
is given by the integration of its RGE, as explained above.  In the
scenario examined in this paper (i.e. supersymmetric see-saw), and  in
the leading-log approximation, ${m_L^2}_{ij}\propto  ({\bf
Y_\nu^+}{\bf Y_\nu})_{ij}$, as given by eq.(\ref{softafterRG}), so
\bea
\label{BR3}
{\mathrm{BR}}(l_i\rightarrow l_j, \gamma) \sim
\frac{\alpha^3}{G_F^2m_S^8}
\left|\frac{-1}{8\pi^2}(3m_0^2+A_0^2)\log\frac{M_X}{M}\right|^2
\left|({\bf Y_\nu^+}{\bf Y_\nu})_{ij}\right|^2\tan^2\beta 
\eea
The ${\bf Y_\nu^+}{\bf Y_\nu}$ matrix, which is therefore the crucial
quantity, has been evaluated in sect.~2 in the general and in
particular cases. Next, we apply those results to the computation of
BR($l_i\rightarrow l_j, \gamma$),

%
%
%
%
%

\section{Experimental data on neutrinos and predictions for 
BR($l_i\rightarrow l_j, \gamma$)}

We study here the expectable predictions for
$l_i\rightarrow l_j, \gamma$ from the existing experimental
information on neutrinos. Let us mention that the present and
forthcoming experimental bounds on $l_i\rightarrow l_j, \gamma$  make
$\mu\rightarrow e, \gamma$ the process with higher potential to
constrain theories involving flavour violation. The present limit is
BR($\mu\rightarrow e, \gamma$) $<$ $1.2\times 10^{-11}$ 
\cite{Brooks:1999pu} and it is going to improve in the near 
future (year 2003) until $\sim 10^{-14}$ \cite{PSImeg}.
  Hence,
we will pay special attention to $\mu\rightarrow e,
\gamma$. Nevertheless, as we will see,  other processes like
$\tau\rightarrow \mu, \gamma$ and  $\tau\rightarrow e, \gamma$ offer
information complementary to  that from $\mu\rightarrow e, \gamma$,
which in some cases can be extremely important.  For $\tau\rightarrow
\mu, \gamma$, in particular, the present (future)  upper limit is $1.1
\times 10^{-6}$ ($\sim 10^{-9}$) \cite{Ahmed:2000gh,private}. 
Finally, there exist
interesting  limits on other branching ratios, like BR($\mu\rightarrow
eee$) $<$ $10^{-12}$ and BR($\mu\ Ti\rightarrow e\ Ti$) $<$ $6\times
10^{-13}$ \cite{Bellgardt,Wintz}. The latter is also to improve in projected
experiments \cite{bachmann}.
These limits are certainly very small, 
but the theoretical predictions  on the branching ratios are also much smaller
due to the presence of extra electromagnetic vertices \cite{kunoreview}

From the results of the previous section, namely eq.(\ref{BR3}), we
know that ${\mathrm{BR}}(l_i\rightarrow l_j, \gamma)\propto
\left|({\bf Y_\nu^+}{\bf Y_\nu})_{ij}\right|^2$. On the other hand,
from   the general equation (\ref{Ynu+Ynu}), it is clear that the
matrix element  $({\bf Y_\nu^+}{\bf Y_\nu})_{ij}$ will depend on
several facts: the low-energy spectrum of neutrinos (contained in
$D_{\kappa}$), the neutrino mixing angles (contained in $U$), the
right-handed neutrino spectrum (contained in $D_{\cal M}$) and the
choice of the $R$ matrix. Let us discuss them in order.

The experimental (solar and atmospheric) data \cite{SK} strongly suggest a
hierarchy of neutrino mass-splittings, $\Delta \kappa_{sol}^2\ll
\Delta \kappa_{atm}^2$. Numerically, $\Delta \kappa_{atm}^2 \simeq
(1.4-6.1)\times 10^{-3} {\mathrm eV}^2/v_2^4$ \cite{Gonzalez-Garcia:2001sq}  
(with central
value around $3\times 10^{-3} {\mathrm eV}^2/v_2^4$),  while the value
of $\Delta \kappa_{sol}^2$ depends on the solution considered to
explain the solar neutrino problem \cite{Pontecorvo,MSW}. The most 
favoured one from the
recent analyses of data \cite{Gonzalez-Garcia:2001sq} 
is the large-angle MSW solution
(LAMSW),  which requires $\Delta \kappa_{sol}^2\sim 3 \times10^{-5}
{\mathrm eV}^2/v_2^4$.  Other solutions are (in order of reliability
from these analyses) the LOW MSW-solution, the small angle
MSW-solution (SAMSW) and the vacuum oscillations solution (VO). They
require, in ${\mathrm eV}^2/v_2^4$ units,  $\Delta \kappa_{sol}^2\sim
10^{-7}$, $5\times 10^{-6}$ and  $8\times 10^{-10}$ respectively. The
corresponding uncertainties are discussed e.g. 
in ref.\cite{Gonzalez-Garcia:2001sq}.  In any
case, there are basically three types of neutrino spectra consistent
with the hierarchy of mass-splittings \cite{altfer}: {\em hierarchical}
($\kappa_1^2\ll\kappa_2^2\ll\kappa_3^2$), {\em ``intermediate''}
($\kappa_1^2\sim\kappa_2^2\gg\kappa_3^2$) and {\em ``degenerate''}
($\kappa_1^2\sim\kappa_2^2\sim\kappa_3^2$). We follow the usual
notation  in which $\kappa_3$ corresponds to the most split mass
eigenvalue, so that $\Delta \kappa_{atm}^2\equiv \Delta
\kappa_{32}^2$, $\Delta \kappa_{sol}^2\equiv \Delta \kappa_{21}^2$.

The hierarchical scenario seems perhaps the most  natural one, since
it resembles the other fermion spectra.  Indeed, it is the natural
spectrum in GUT theories, particularly $SO(10)$ models.  In any case,
it is the most extensively analyzed scenario in the literature and we
will pay special attention to it.  In this scenario one should take
$\kappa_3^2 \simeq \Delta \kappa_{atm}^2$,  $\kappa_2^2 \simeq \Delta
\kappa_{sol}^2$, $\kappa_1^2 \simeq 0$.  The other two scenarios
(intermediate and degenerate) are also interesting and we will address
them. In particular, the degenerate scenario is the only one
compatible with a  relevant cosmological role of neutrinos (provided
$\kappa\simeq {\cal O}(1)  {\mathrm eV}/v_2^2$).

Concerning the mixing angles, $\theta_{23}$ and $\theta_{13}$ are
constrained by the atmospheric and CHOOZ data to be near maximal and
minimal, respectively.  The $\theta_{12}$ angle depends on the solution
considered to explain the solar neutrino problem.  In the LAMSW case,
$\theta_{12}$ is near maximal. In the other explanations of the solar
neutrino problem $\theta_{12}$ can be either minimal (SAMSW), or near
maximal (VO and LOW). The precise preferred values for the angles
depend on the sets of data used in the fits. For example, from the
global analysis of ref.\cite{Gonzalez-Garcia:2001sq} 
these are $\tan^2\theta_{13}\sim 0.005$,
$\tan^2\theta_{23}\sim 1.4$ and, for the LAMSW, $\tan^2\theta_{12}\sim
0.36$. The uncertainties depend also on the type of fit, as well as on
the values used for the mass splittings.  In summary, the two basic
forms that $U$ can present are either a single-maximal or (more
plausibly) a bimaximal mixing matrix. Schematically,
\be
\label{Uaprox} 
U\sim\pmatrix{1 & 0 & 0\cr 0& \frac{1}{\sqrt{2}} &
\frac{1}{\sqrt{2}}\cr 0 & -\frac{1}{\sqrt{2}} & \frac{1}{\sqrt{2}} } \
\ \ \ {\mathrm or} \ \ \ \ U\sim\pmatrix{\frac{1}{\sqrt{2}} &
\frac{1}{\sqrt{2}} & 0\cr -\frac{1}{2}& \frac{1}{2}&
\frac{1}{\sqrt{2}}\cr \frac{1}{2} & -\frac{1}{2} & \frac{1}{\sqrt{2}}
}\ , \ee

With regard to the spectrum of right-handed neutrinos, there are two basic
possibilities: either they are hierarchical, i.e. ${\cal M}_1\ll{\cal
M}_2\ll{\cal M}_3$, or they are degenerate, i.e. ${\cal M}_1\sim{\cal
M}_2\sim{\cal M}_3$.  Both of them were considered in sect.~2, where it
was also shown that the first possibility counts the case  of an
``inverse hierarchy'' or any other ordering of the ${\cal M}$
eigenvalues.


Finally, $R$ is the generic (complex) orthogonal matrix discussed in
sect.~2.  In principle, it contains three independent complex
parameters, but we saw that this number can be usually reduced in  a
drastic way.

Let us analyze now the predictions for BR($l_i\rightarrow l_j,
\gamma$) in the different scenarios above exposed.

\subsection{$\nu_L$'s and $\nu_R$'s completely hierarchical}    

\subsubsection{The generic case}

We consider first the case where both the left-handed and the
right-handed neutrinos are completely hierarchical, i.e.
\be
\label{jerar2}
D_\kappa\simeq {\mathrm diag}(0,\kappa_2,\kappa_3),\;\;\;\;  D_{\cal
{M}}\simeq {\mathrm diag}(0,0,{\cal M}_3)\; .    \ee
We are neglecting for the moment contributions of ${\cal O}({\cal
M}_2)$. (Later on, we will discuss when those contributions would be
relevant.)  The general form of the ${\bf Y_\nu}$ matrix in this case
was already worked out in subsect.~2.1. If  $R_{32}\neq 0$ or $R_{33}\neq
0$ (later on we will consider the case $R_{32},R_{33} \simeq 0$),
${\bf Y_\nu}$ is given by eq.(\ref{Ynujer}), so only the last row of
${\bf Y_\nu}$ has sizeable entries. Thus, the relevant quantity $({\bf
Y_\nu^+}{\bf Y_\nu})_{ij}$ has the  form
\bea
\label{Y+Yjer}
({\bf Y_\nu^+}{\bf Y_\nu})_{ij}=({\bf Y_\nu})_{3i}^*({\bf
Y_\nu})_{3j}= {\cal M}_3 \left[\sum_{l=2,3}
R_{3l}^*\sqrt{\kappa_l}U_{il}\right] \left[\sum_{l'=2,3}
R_{3l}\sqrt{\kappa_l'}U_{jl'}^*\right]  \eea
%
%
Focusing on $\mu\rightarrow e, \gamma$, and using the parametrization
of $R$  of eq.(\ref{R}), eq.(\ref{BR3}) reads
\bea
\label{BR4}
{\mathrm{BR}}(\mu\rightarrow e, \gamma) &\propto&   \left|({\bf
Y_\nu^+}{\bf Y_\nu})_{21}\right|^2 \nonumber \\  &=& \left|\ {\cal
M}_3 |\hat c_2|^2 \left[\hat
s_1^*\sqrt{\kappa_2}U_{22}+\hat c_1^*\sqrt{\kappa_3}U_{23}\right] 
\left[\hat s_1\sqrt{\kappa_2}U_{12}^*+\hat
c_1\sqrt{\kappa_3}U_{13}^*\right] \
\right|^2   \eea
Let us recall here that, since $\kappa_2$ and $\kappa_3$ represent the
solar and atmospheric splitting, respectively, then
$\kappa_2\ll\kappa_3$. Also, the matrix element $U_{13}\sim s_{13}\ll 1$,
as $\theta_{13}$ is the mixing angle constrained by CHOOZ.  In the SAMSW
scenario, $U_{12}$ and $U_{23}$ are also extremely small  (see
eq.(\ref{Vdef})), so $\mu\rightarrow e,\gamma$ is
suppressed. However, in the other scenarios (LAMSW, VO and LOW) $U$
is basically a bimaximal mixing matrix. Then  $U_{12}, U_{23}\simeq
1/\sqrt{2}$,  $U_{22}\simeq{1/2}$. Let us recall that this precisely
includes the  scenarios favoured from present data, i.e. LAMSW and LOW.

The prefactor ${\cal M}_3 |\hat c_2|^2$ in eq.(\ref{BR3})  can be
written as
\bea
\label{unifcond}
{\cal M}_3 |\hat c_2|^2 = \frac{|{Y}_0|^2}{|\hat s_1|^2\kappa_2+ |\hat
c_1|^2\kappa_3} \eea
where $|{Y}_0|^2$ is the largest eigenvalue of $({\bf Y_\nu^+}{\bf
Y_\nu})$. So, the right hand side of eq.(\ref{BR4}) depends just on
two independent parameters, $|{Y}_0|$ and $\hat \theta_1$ (the
remaining parameters have a precise physical meaning and are
measurable at low energy). If we further impose a unification
condition, like $|{Y}_0(M_X)|=|Y_t(M_X)|$,  BR($\mu\rightarrow e,
\gamma$) depends just on one parameter, $\hat \theta_1$.  This
conclusion could also be obtained from eq.(\ref{Ynujer2}).

Now, as discussed in sect.~2, the results of ref.
\cite{Hisano:1999fj} correspond to
take $R={\bf 1}$, i.e. $\hat c_i=1$. Then, BR($\mu\rightarrow e,
\gamma$), as written in eq.(\ref{BR4}) is suppressed by  $|U_{13}|^2
\ll 1$. The authors of ref.\cite{Hisano:1999fj}
 considered however the contributions
%
%
%
%
of ${\cal O}({\cal M}_2)$ (which we have not considered yet). Then,
working in the LAMSW scenario, they obtained a non-vanishing result
for BR($\mu\rightarrow e, \gamma$), as a function of ${\cal M}_2$.
What is apparent from eqs.(\ref{BR4}, \ref{unifcond})  is that that
result corresponds to a very special point in the parameter space,
namely $\hat c_1=1$. In general, $\hat c_1\neq 1, \hat s_1\neq 0$, so
the prediction for  BR($\mu\rightarrow e, \gamma$) is much larger.
For random values of $\hat \theta_1$  we expect
\bea
\label{Y+Yaprox} 
({\bf Y_\nu^+}{\bf Y_\nu})_{21}\sim \frac{|{Y}_0|^2}{|\hat
s_1|^2\kappa_2+ |\hat c_1|^2\kappa_3} \hat
c_1^*\hat s_1  \sqrt{\kappa_3 \kappa_2}U_{23} U_{12}^*  \eea
This formula allows to compare the foreseeable predictions for
BR($\mu\rightarrow e, \gamma$) in different scenarios, taking into
account the different values for $\kappa_2$, $U_{ij}$. Clearly, the
largest values for BR($\mu\rightarrow e, \gamma$) are  obtained for
the LAMSW scenario (precisely, the most favoured one by  the present
analyses). For the LOW and SAMSW scenarios one expects predictions
$\sim 20$ and $\sim 1000$ times smaller respectively.  Indeed, for the
LAMSW scenario the previous equation generically gives a branching
ratio above the {\em present} experimental limits,  at least for
${Y}_0 = {\cal O}(1)$, as it occurs in the unified scenarios!

\begin{figure}
\vspace{-0.3cm}
\centerline{\vbox{
\psfig{figure=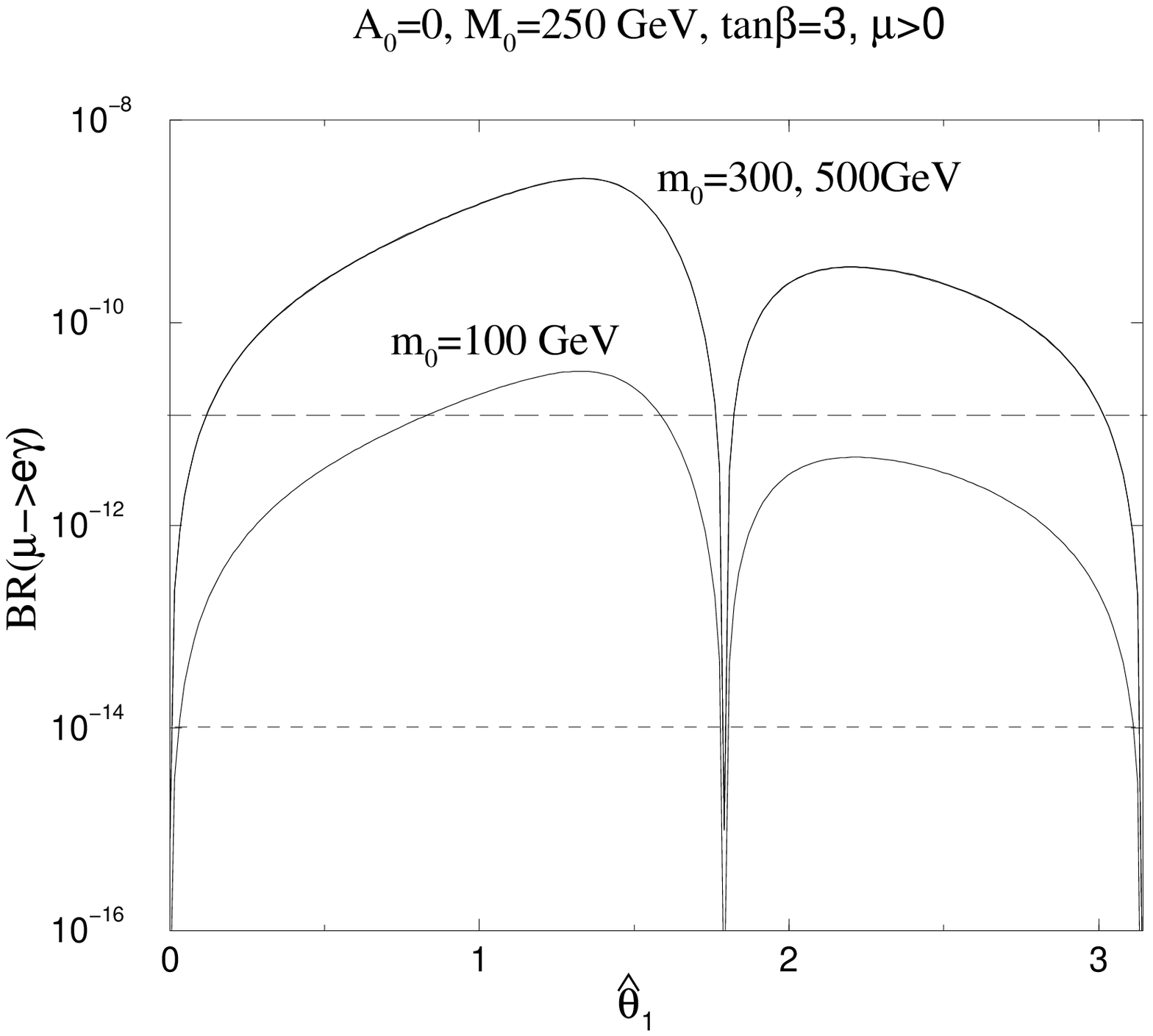,height=10cm}
\psfig{figure=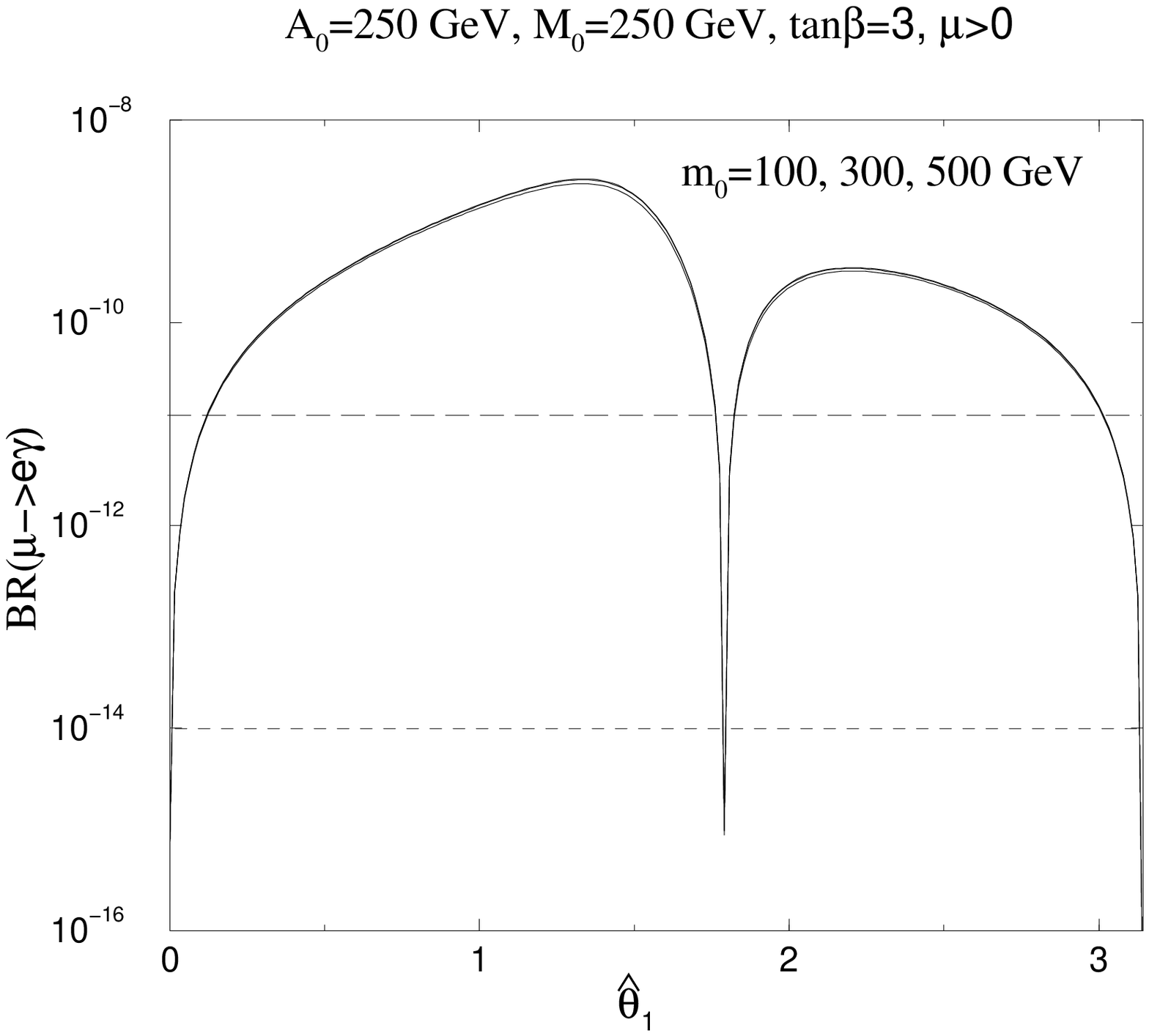,height=10cm}}}
\caption
{\footnotesize  
Branching ratio of the process $\mu\rightarrow e, \gamma$
vs. the unknown angle $\hat \theta_1$ for the case of hierarchical (left and right) neutrinos and
two typical sets of 
supersymmetric parameters, as indicated in the plots. The dashed lines 
correspond to the present and forthcoming upper bounds. 
A top-neutrino ``unification'' condition has been used to fix
the value of the largest neutrino Yukawa coupling at high energy.
$\hat \theta_1$ is taken real for simplicity, so the two limits
$\hat \theta_1=0,\pi$ of the horizontal axis represent the same
physical point.
In the upper plot the curves corresponding to $m_0=300$ GeV and
$m_0=500$ GeV are almost indistinguishable. The same is true in 
the lower plot for the curves corresponding to $m_0=100$ GeV, 
$m_0=300$ GeV and $m_0=500$ GeV.
}
\end{figure}

\begin{figure}
\centerline{\vbox{
\psfig{figure=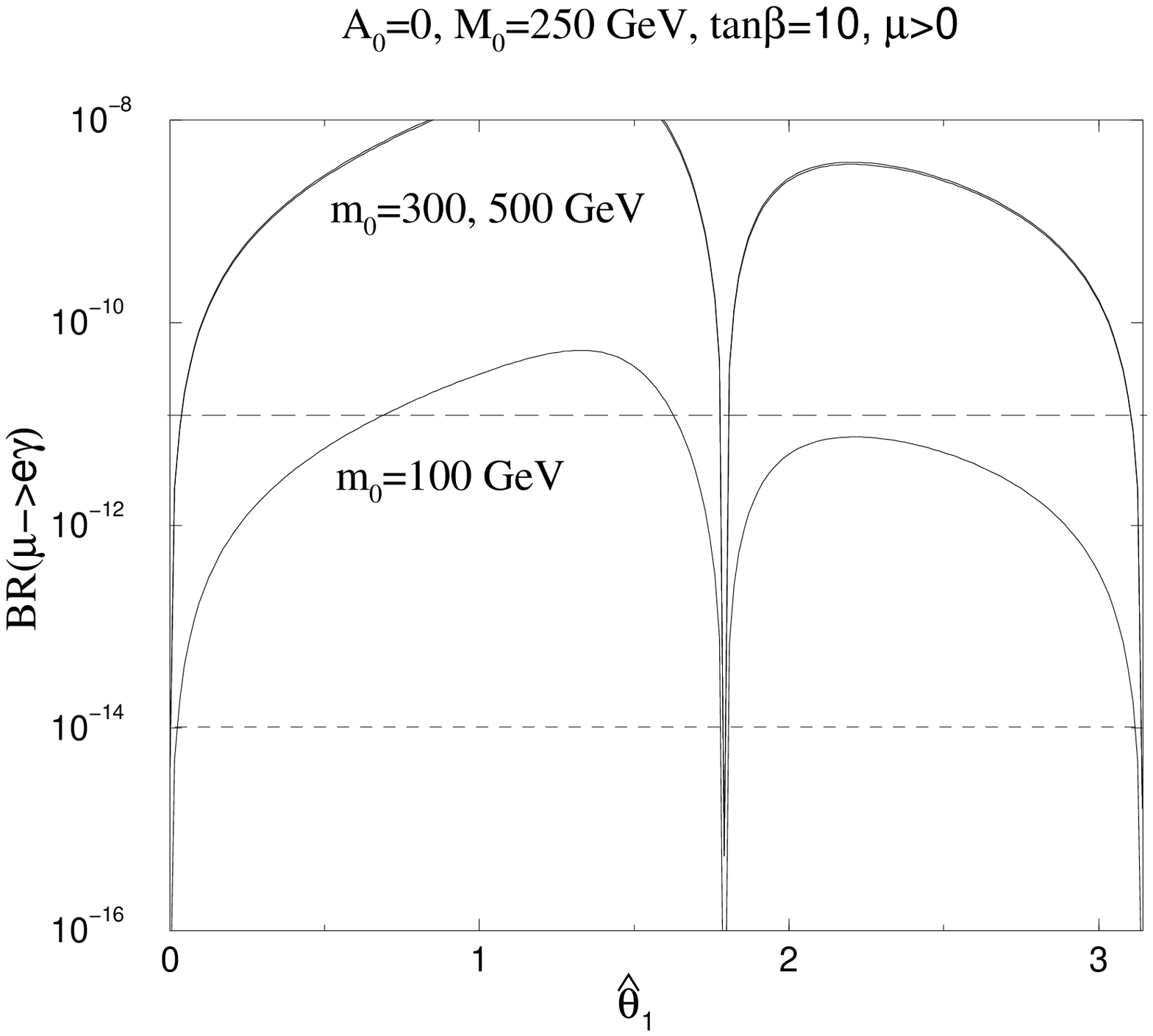,height=10cm}
\psfig{figure=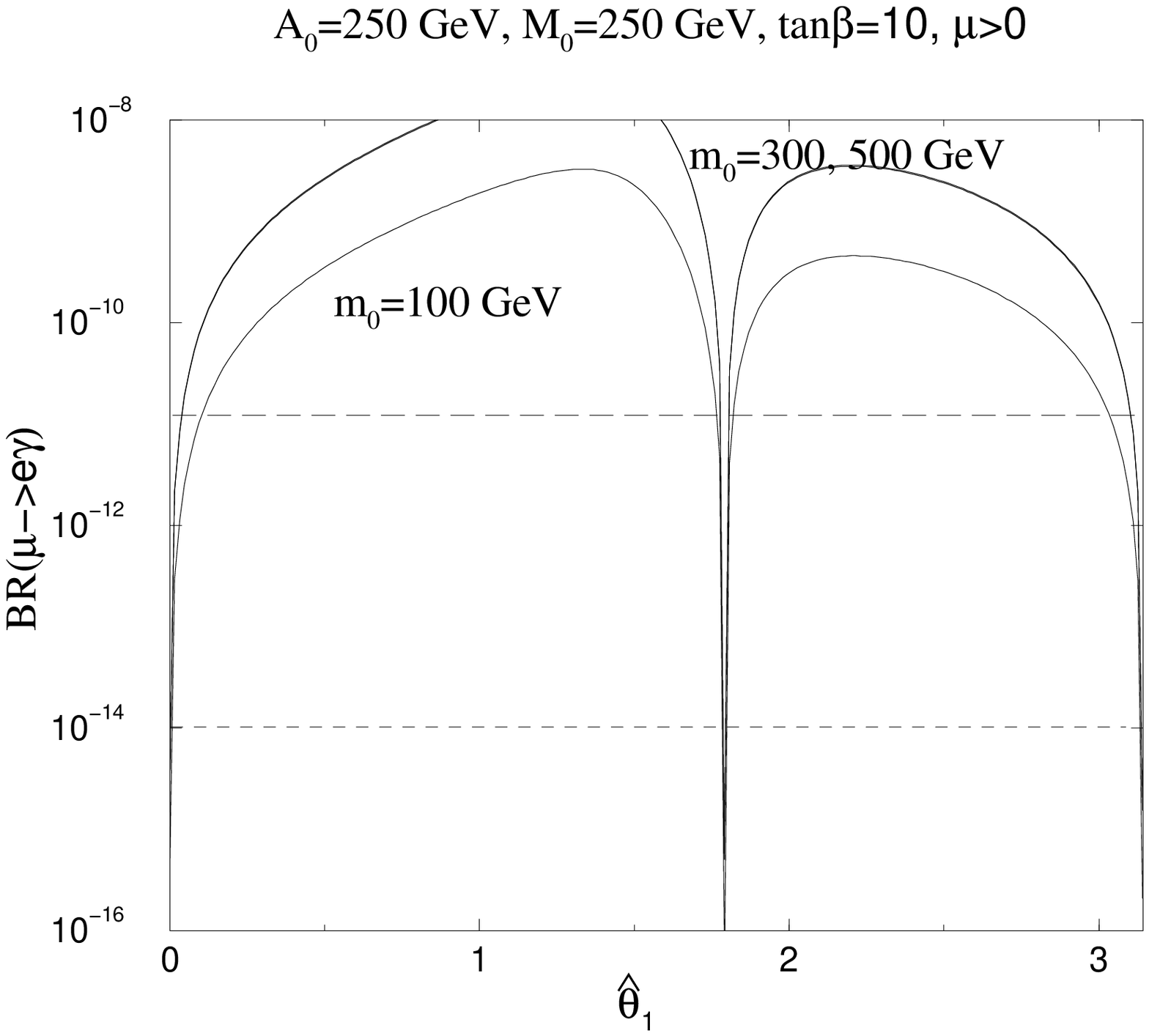,height=10cm}}}
\caption
{\footnotesize  The same as Fig.~3, but for $\tan \beta=10$.
}
\end{figure}

This is apparent from the plots of Fig.~3, where it is shown
BR($\mu\rightarrow e, \gamma$) vs. $\hat \theta_1$ in the LAMSW
scenario for different typical sets of supersymmetric parameters,
using a unification condition for the largest neutrino Yukawa
coupling, i.e. $|Y_0(M_X)|^2 = |Y_t(M_X)|^2$, with $M_X=2\times
10^{16}$ GeV.  Incidentally, the value of the supersymmetric $\mu$-parameter is extracted (here and throughout the paper) from the electroweak breaking condition, choosing $\mu>0$. We have not shown the results for $\mu<0$ since they are always very similar. From Fig.~3, it is clear that most of the
parameter space (i.e. the range of $\hat \theta_1$) is already
excluded by the present bounds on BR($\mu\rightarrow e, \gamma$),  or
it will become probed by the next generation of $\mu\rightarrow e,
\gamma$ experiments (PSI), scheduled for 2003. For the sake of
clarity, we took $\hat \theta_1$ real in Fig.~3. The results for $\hat
\theta_1$ complex are very similar, as can be seen e.g. from
eq.(\ref{Y+Yaprox}). Fig.~4 is completely analogous to Fig.~3, but for
$\tan \beta=10$ (instead of $\tan \beta=3$). It illustrates the strong 
dependence of  BR($\mu\rightarrow e, \gamma$) on $\tan \beta$ in
 the expected way ($\sim \tan^2 \beta$).  An
alternative representation of the allowed parameter space is given in
Fig.5, where $\hat \theta_1$ is given in logarithmic units (the set of
supersymmetric parameters corresponds to that of the second plot of
Fig.3). These can be more meaningful in particular model-building
constructions. Of course, the basic results remain: $\hat \theta_1$
should be $< 0.12$ ($< 3 \times 10^{-3}$) with the present (future) upper
bounds on BR($\mu\rightarrow e, \gamma$). This corresponds to  ${\bf
Y}_{31} < 0.04 {\bf Y}_{33}$ 
 (${\bf Y}_{31} < 9 \times 10^{-4} {\bf Y}_{33}$).
  This gives a measure of the constraints  on the
textures. (In a logarithmic plot it is not possible to say what
percent  of the parameter space is allowed since this includes the
origin.)

%
%
\begin{figure}
\centerline{\vbox{
\psfig{figure=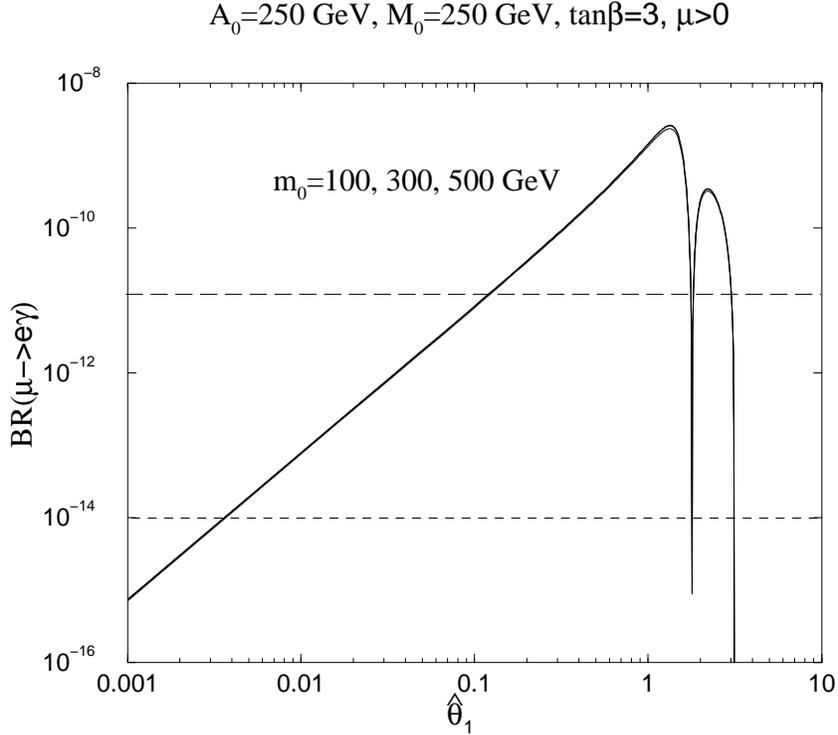,height=10cm}}}
\caption
{\footnotesize  
The same as Fig.3 (lower plot) using a logarithmic scale for $\hat \theta_1$.
}
\end{figure}
%
It is
worth-noticing from the curves of Figs. 3, 4, the existence of two values  
of $\hat
\theta_1$, namely $\hat \theta_1=0$, $\hat \theta_1\sim 2$, in whose
neighborhood the value of BR($\mu\rightarrow e, \gamma$) is
drastically suppressed. We will comment shortly on this feature. For
large enough values of the universal soft scalar mass, $m_0$,
BR($\mu\rightarrow e, \gamma$) decreases, as expected: for large $m_0$, 
the branching ratio goes as $\sim m_0^{-4}$, as it is apparent from Fig.~6.
Typically, for $\tan\beta=3$ BR($\mu\rightarrow e, \gamma$) does not
fall below the present or forthcoming experimental upper bounds until
$m_0\simgt 1.5$ TeV, i.e. beyond the reasonable range to avoid
fine-tuning problems. For larger $\tan\beta$, this extremal value increases, 
since BR($\mu\rightarrow e, \gamma$) $\sim$ $\tan^2 \beta/m_0^{4}$. Notice 
that there is a very narrow window at around 120 GeV, for $A_0=0$ GeV, where 
the branching ratio is very suppressed. This happens for every value of
$\hat \theta_1$ and is the result of a cancellation between the chargino 
and the neutralino amplitudes.
%
%
\begin{figure}
\centerline{\vbox{
\psfig{figure=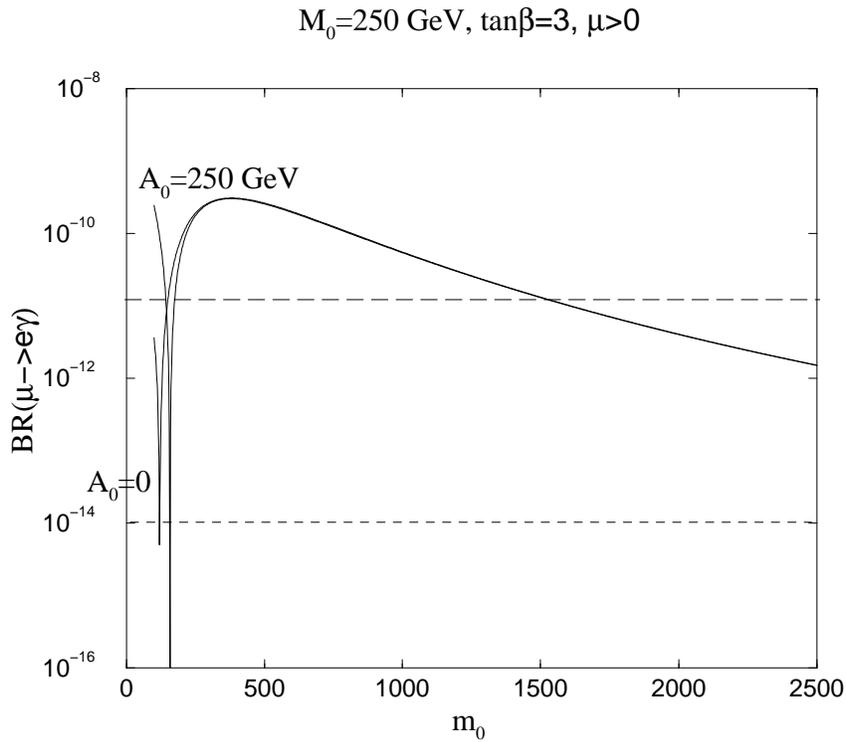,height=10cm}}}
\caption
{\footnotesize  
Branching ratio of the process $\mu\rightarrow e, \gamma$ 
vs. the universal scalar mass, $m_0$, in the case of hierarchical 
(left and right) neutrinos, and for the two 
sets of supersymmetric parameters of Fig.~3 (with $\tan\beta=3$)
and $\hat \theta_1=0.5$. 
The dashed lines correspond to the present and forthcoming upper bounds.
Except for a narrow range at low energy, the prediction does not fall 
below the present upper bound until $m_0 > 1.5$ TeV. The same occurs 
for other values of $\hat \theta_1$, except for the two special values 
seeable in Fig.~3 and discussed in subsect.~4.1.2. 
}
\end{figure}
%

Finally, if one relaxes the unification condition on $|Y_0(M_X)|$,
BR($\mu\rightarrow e, \gamma$) logically decreases, as shown in
Fig.~7. Anyway, it is noticeable that even for $Y_0(M_X)$ ten times
smaller than  $Y_t(M_X)$, the conclusion remains that most of the
parameter space will be probed in the forthcoming experiments.


\begin{figure}
\centerline{\vbox{
\psfig{figure=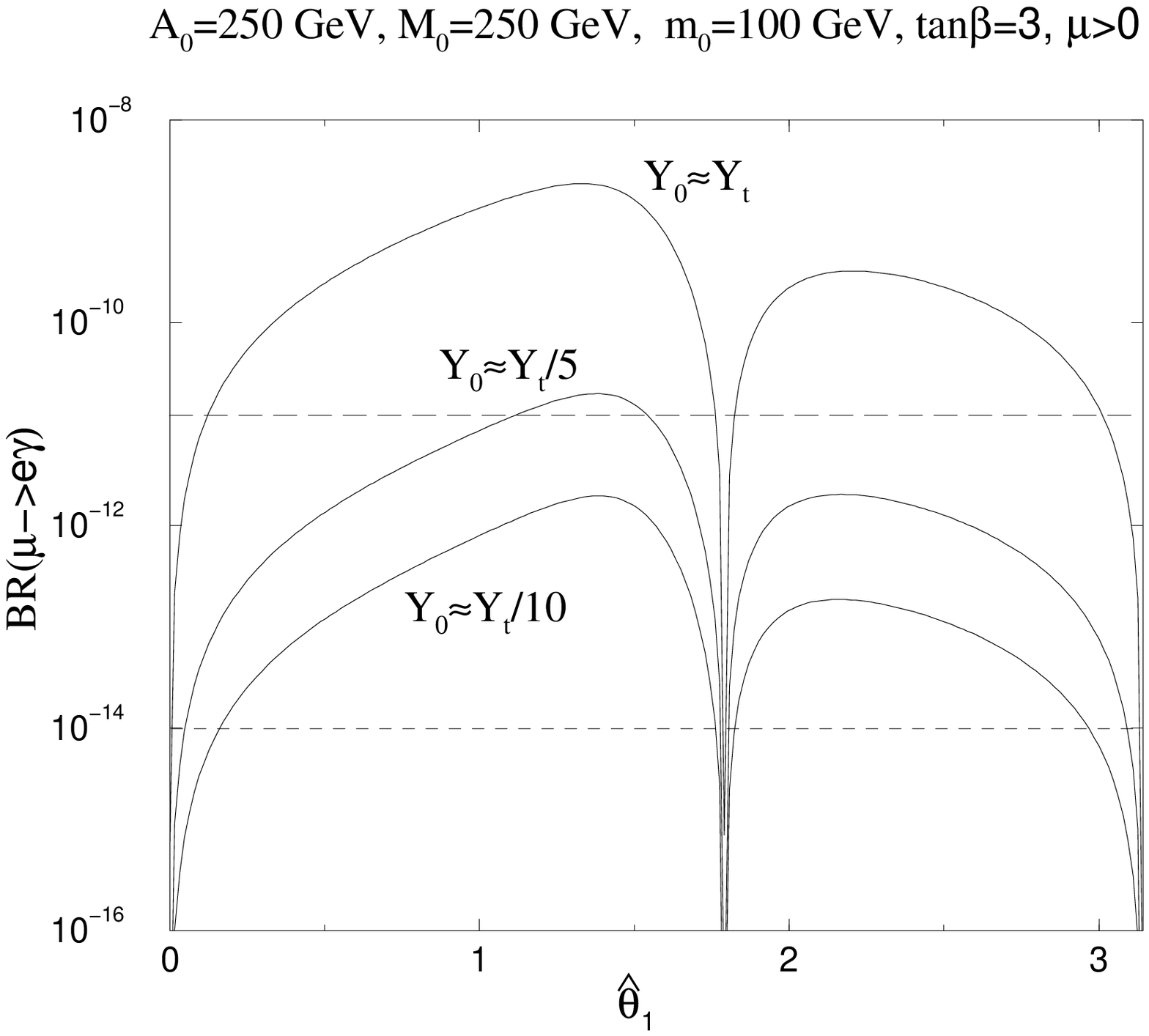,height=10cm}}}
\caption
{\footnotesize  
The same as Fig.~3 (lower plot) for different values of the largest 
neutrino Yukawa coupling, $Y_0$, at the ``unification'' scale, 
$M_X$. $Y_t$ denotes the value of the top Yukawa coupling at that 
scale.
}
\end{figure}

\subsubsection{Special textures}

Let us now turn to the two values of $\hat \theta_1$ where
BR($\mu\rightarrow e, \gamma$) is suppressed. They correspond to the
values for which $({\bf Y_\nu^+}{\bf Y_\nu})_{21}$, as given by
eq.(\ref{Y+Yjer}), is small.  Notice that there are two possible ways
to fulfill this requirement, namely $({\bf Y_\nu})_{31}\simeq 0$ or
$({\bf Y_\nu})_{32}\simeq 0$, which translate into
\bea
\tan \hat \theta_1 &\simeq&
-\sqrt{\frac{\kappa_3}{\kappa_2}}\frac{U_{13}^*}{U_{12}^*} =
-\sqrt{\frac{\kappa_3}{\kappa_2}}\frac{V_{13}^*}{V_{12}^*} \simeq 0
\label{puntos1}
\\ \tan \hat \theta_1 &\simeq&
-\sqrt{\frac{\kappa_3}{\kappa_2}}\frac{U_{23}^*}{U_{22}^*}  =
-\sqrt{\frac{\kappa_3}{\kappa_2}}\frac{V_{23}^*}{V_{22}^*}\;\;,
\label{puntos2}
\eea
where we have used eqs.(\ref{Ynujer2}, \ref{UV}).  We do not know any
reason why any  of the previous conditions should be satisfied in a
particular model, although of course it is not excluded.  To this
respect, the first condition, i.e. $\hat \theta_1\simeq 0$, seems to
be more ``natural'' to obtain than the second one: note that $\hat
\theta_i= 0$, i.e. $R={\bf 1}$, corresponds to the case where all the
mixing can be attributed to the sector of charged leptons, as
discussed in sect.~2, which is a physical possibility.
It is interesting to explicitly show the textures of ${\bf Y_\nu}$
that correspond to these `privileged' values of  $\hat \theta_1$ from
eqs.(\ref{puntos1}, \ref{puntos2}). They read, respectively,
\bea  {\bf Y_\nu}&\propto&\pmatrix{0 & 0 & 0\cr 0& 0& 0\cr 0 & V_{31}
&  -V_{21} }  \propto \pmatrix{0 & 0 & 0\cr 0& 0& 0\cr 0 & 1 & 1 } \ \
\ \ ,
\label{textures1} 
\\ {\bf Y_\nu}&\propto&\pmatrix{0 & 0& 0\cr 0& 0& 0\cr  -V_{31} & 0 &
V_{11} } \propto \pmatrix{0 & 0& 0\cr 0& 0& 0\cr -1 & 0 & {\sqrt{2}} }
\ \ \ \ .
\label{textures2} 
\eea
Here we have used ${\mathrm Im}V_{11}=0$ and $s_{13}\ll 1$, so that 
${\mathrm Im}V_{21},{\mathrm Im}V_{31}\ll 1$. 
Besides,
the numerical values of the entries in (\ref{textures1}, \ref{textures2}) 
correspond to a  bimaximal mixing
$V$ matrix, which includes the preferred LAMSW scenario.  Let us
stress that the previous textures are to be understood in the basis we
have chosen to work, i.e. where both ${\bf Y_e}$ and ${\cal M}$ are
diagonal, and the latter has the eigenvalues ordered in the usual
hierarchical way shown in eq.(\ref{jerar2}).
It is interesting to note that texture (\ref{textures1}) has been 
advocated in ref.\cite{altfer} on different grounds.

Some comments are in order here. As was discussed in subsect.~2.1, the
first two rows of ${\bf Y_\nu}$, though suppressed, are normally
important to reconstruct $\kappa$ from ${\bf Y_\nu}$, ${\cal M}$. In
particular, for texture (\ref{textures2}) the precise values of
the first two rows are crucial to reproduce the bimaximal structure of
$\kappa$. On the other hand, this is not necessarily the case for
texture (\ref{textures1}). Actually, it is remarkable that texture 
(\ref{textures1}) works
fine, both to reproduce $\kappa$ and to give suppressed
${\mathrm{BR}}(\mu\rightarrow e, \gamma)$, even if it is given in a
basis where  ${\cal M}$ is not diagonal.

In general, it is clear from eq.(\ref{Y+Yjer}) that the
 $l_i\rightarrow l_j, \gamma$ process is suppressed at the points in
 the parameter space where $({\bf Y_\nu})_{3i}\simeq 0$ or $({\bf
 Y_\nu})_{3j}\simeq 0$. Therefore, there are three special textures:
 the two given in the above eqs.(\ref{textures1}, \ref{textures2}), which
 correspond to  $({\bf Y_\nu})_{31}\simeq 0$ and $({\bf
 Y_\nu})_{32}\simeq 0$, and the one corresponding to $({\bf
 Y_\nu})_{33}\simeq 0$. The latter implies
\bea \tan \hat \theta_1 \simeq
-\sqrt{\frac{\kappa_3}{\kappa_2}}\frac{U_{33}^*}{U_{32}^*}  =
-\sqrt{\frac{\kappa_3}{\kappa_2}}\frac{V_{33}^*}{V_{32}^*}\;\;,
\label{puntos3}
\eea
and the associated texture is
\bea
\label{textures3} 
{\bf Y_\nu}\propto\pmatrix{0 & 0 & 0\cr 0& 0& 0\cr V_{21} & -V_{11} &
0 }  \propto \pmatrix{0 & 0 & 0\cr 0& 0& 0\cr -1 & \sqrt{2} & 0 } \ \
\ \ , \eea
where, again, the numerical values values correspond to the bimaximal
scenario.  If the texture of ${\bf Y_\nu}$ is one of the special forms
(\ref{textures1}, \ref{textures2}, \ref{textures3}), there are two
$l_i\rightarrow l_j, \gamma$ processes suppressed and one
unsuppressed.  This means that for those privileged cases where
$\mu\rightarrow e, \gamma$ is unusually suppressed, the
$\tau\rightarrow \mu, \gamma$ and  $\tau\rightarrow e, \gamma$
processes are relevant and may give the most important restrictions on
models. In particular, if ${\bf Y_\nu}$ is as in  eq.(\ref{textures1})
or eq.(\ref{textures2}) (i.e. $({\bf Y_\nu})_{31}\simeq 0$  or $({\bf
Y_\nu})_{32}\simeq 0$), then  ${\mathrm{BR}}(\tau\rightarrow \mu,
\gamma)$ or ${\mathrm{BR}}(\tau\rightarrow e, \gamma)$ respectively
are unsuppressed.

%

We have illustrated this interesting aspect by plotting
BR($\tau\rightarrow \mu, \gamma$) vs. $\hat \theta_1$  in Fig.~8 for
one of the typical supersymmetric scenarios considered in
Fig.~3. Notice that for the special value $\hat \theta_1\simeq 0$, at
which ${\mathrm{BR}}(\mu\rightarrow e, \gamma)$ is suppressed,
BR($\tau\rightarrow \mu, \gamma$) is not, lying above the  forthcoming
experimental upper bound.

\begin{figure}
\centerline{\vbox{
\psfig{figure=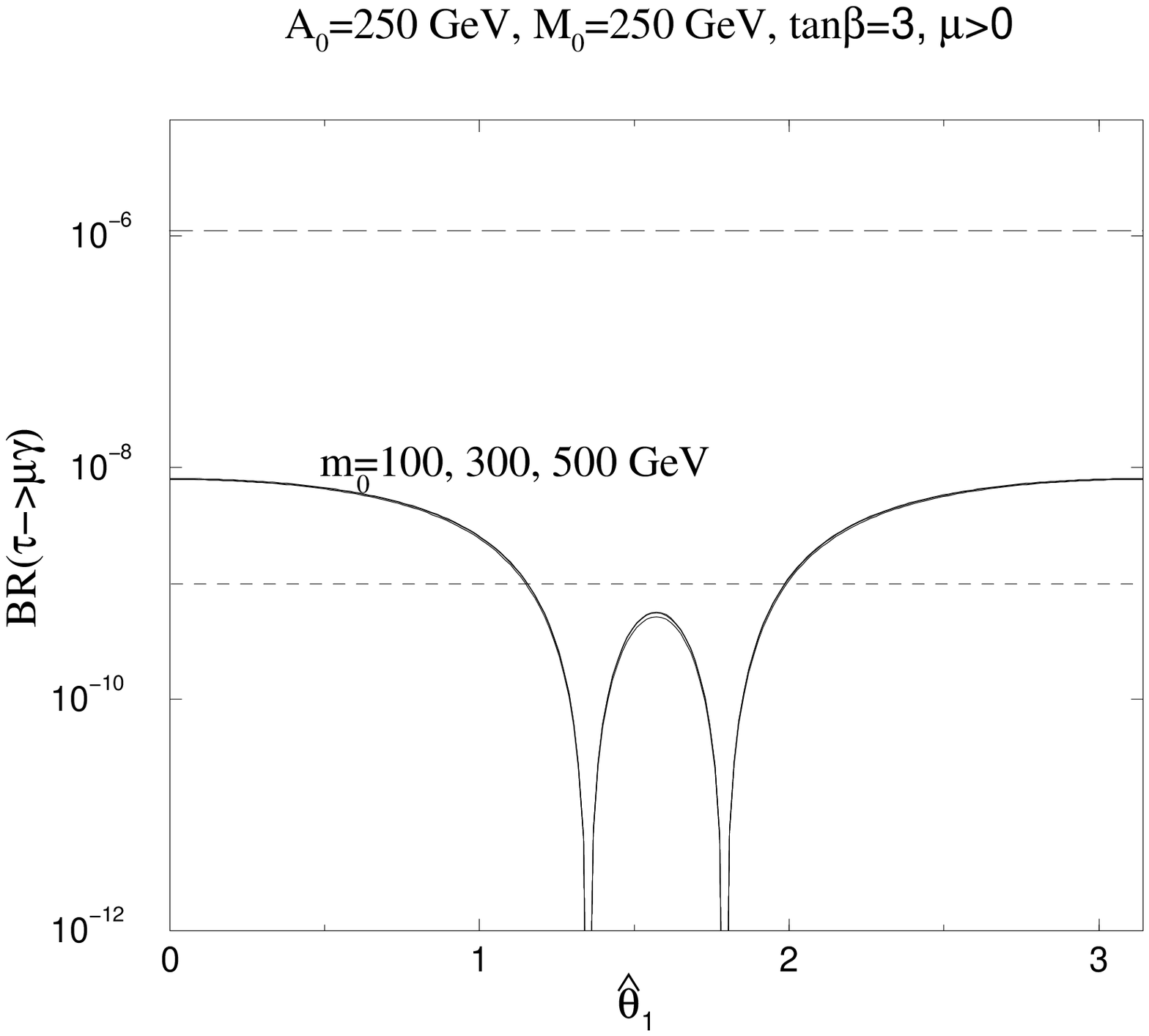,height=10cm}}}
\caption
{\footnotesize  
The same as Fig.~3 (lower plot), but for the branching ratio 
of the process $\tau\rightarrow \mu, \gamma$. The dashed lines 
correspond to the present and forthcoming upper bounds.
}
\end{figure}

\subsubsection{The contribution $\propto {\cal M}_2$}

So far, all the results of this subsection have been obtained
neglecting the value of ${\cal M}_2$ (see eq.(\ref{jerar2})).  Let us
consider here a value of ${\cal M}_2\neq 0$, though smaller  than
${\cal M}_3$, as it was done in ref.\cite{Hisano:1999fj}.

Clearly, for generic values $\hat \theta_i$, the contributions already
computed  [see eqs.(\ref{BR4}--\ref{Y+Yaprox})] will be the dominant
ones. However, if $\hat \theta_1$ corresponds to one of the particular
values of eqs.(\ref{puntos1}, \ref{puntos2}), the contributions $\sim
{\cal M}_2$ can be relevant or even dominant. Normally, this will
happen only for $\hat \theta_1$ quite close to eqs.(\ref{puntos1},
\ref{puntos2}).  Suppose for example that $R\simeq {\bf 1}$,
i.e. $\hat s_i$ small and real.  This corresponds to eq.(\ref{puntos1}) and it
is probably the most interesting case (and the one considered in
ref. \cite{Hisano:1999fj}). Then  $({\bf Y_\nu^+}{\bf Y_\nu})_{ij}
 =  ({\bf
Y_\nu})_{2i}^*({\bf Y_\nu})_{2j} + ({\bf Y_\nu})_{3i}^*({\bf
Y_\nu})_{3j}$ with ${\bf Y_\nu}$ given by eq.(\ref{Ynu}). Keeping only
terms up to ${\cal O}(\hat s_i)$ we get
\bea
\label{Y+Yaprox2}
({\bf Y_\nu^+}{\bf Y_\nu})_{21}&\simeq& {\cal M}_2\kappa_2
U_{12}^*U_{22} + {\cal M}_3\kappa_3 U_{13}^*U_{23} \nonumber \\ &+&
\hat s_1({\cal M}_3-{\cal M}_2)\sqrt{\kappa_2 \kappa_3 } \left(
U_{13}^*U_{22}  +  U_{12}^*U_{23} \right), \eea
so that in a bimaximal case, even taking $U_{13}=0$, the part
proportional to ${\cal M}_3$ will be indeed the dominant one except
for $ \hat s_1 \simlt({\cal M}_2/{\cal
M}_3)(\sqrt{\kappa_2/\kappa_3})$,  i.e. very close to zero.

\vspace{0.3cm} For the specific case $R= {\bf 1}$, i.e. $ \hat s_1 =0$,
 we get  $({\bf
Y_\nu^+}{\bf Y_\nu})_{21}\simeq {\cal M}_2\kappa_2
U_{12}^*U_{22}$. Taking into account from eq.(\ref{unifcond}) $|Y_0|^2
= {\cal M}_3\kappa_3$, the value of  $({\bf Y_\nu^+}{\bf Y_\nu})_{21}$
is exactly as the one we will obtain in another context in
subsect.~4.2, see eq.(\ref{Ynu+Ynu12hierdeg}) below,
but multiplied by ${\cal M}_2/{\cal M}_3$. In consequence, the plots
of ${\mathrm{BR}}(\mu\rightarrow e, \gamma)$ are identical to
those of Fig.~9 below, but  multiplied by $({\cal M}_2/{\cal M}_3)^2$.
For ${\cal M}_2/{\cal M}_3\simgt 10^{-2}$ the branching ratio is  still
testable in the forthcoming generation of experiments.

\subsubsection{$R_{32}\simeq R_{33}\simeq 0$}

So far  we have assumed $R_{32}\neq 0$ or  $R_{33}\neq 0$ in
eq.(\ref{Ynu}). Recall that $R$ is an unknown
orthogonal matrix. The experimental data do not restrict its  form,
although $R$ may be much more definite in specific models.  We
consider now the special case when  $R_{32}\simeq R_{33}\simeq 0$.  It
should be kept in mind that this represents a narrow region of  the
whole parameter space of $R$.

However, as we are about to see, this possibility includes the particular 
case where
all the leptonic flavour violation can be attributed to the sector of
right-handed neutrinos. This happens when there exists a  basis in
which $\bf{Y_e}$ and ${\bf Y_\nu}$ get  simultaneously diagonal, while
${\cal M}$ gets non-diagonal.  This is physically interesting and it
can occur in particular models.  As explained in sect.~2, in the basis
we have chosen to work  ($\bf{Y_e}$ and ${\cal M}$ diagonal), this
implies ${\bf Y_\nu}= W D_Y$, where $D_Y$ is a diagonal matrix and $W$
is a unitary matrix. Hence, ${\bf Y_\nu}^+{\bf Y_\nu}=
D_{|Y^2|}$. Then, from the general equation (\ref{Ynu}) or
(\ref{Ynu+Ynu}), we get
\bea
\label{RDR}
R^+ D_{\cal M} R = D_{ \sqrt{\kappa^{-1}}} U^+ D_{|Y^2|}  U D_{
\sqrt{\kappa^{-1}}}\ .  \eea
If $U$ is of the bimaximal type, as it occurs in the preferred LAMSW
scenario, then $U_{i1}\neq 0$ for any $i$ [see
eq.(\ref{Uaprox})]. This means that, in first approximation, the only
non-vanishing entry of the right hand side of eq.(\ref{RDR}) is the
(1,1), so the same  holds for the left hand side. Since $D_{\cal M}$
is hierarchical,  $(R^+ D_{\cal M} R)_{ij}\simeq R^*_{3i}{\cal M}_3
R_{3j}$. Thus, $R_{32}\simeq R_{33}\simeq 0$ {\em q.e.d.}
A special case occurs
when $D_{|Y^2|} = |Y_0^2|{\bf 1}$. Then eq.(\ref{RDR}) becomes $R^+
D_{\cal M} R = |Y_0^2|D_{ \kappa^{-1}}$ for any form of $U$; and $R$
turns out to be 
simply the  ``permutation matrix'' that interchanges the 1-- and
3--planes: $R=R_p=\left\{(0,0,\pm 1),\ (0,\pm 1,0),\ (\pm 1,0,0)
\right\}$. Of course, in all these cases, since ${\bf Y_\nu}^+{\bf
Y_\nu}$ is diagonal, there is no RG generation of non-diagonal soft
terms, and the predictions for BR($l_i\rightarrow l_j, \gamma$) are as
in the SM, i.e. negligible.

\vspace{0.3cm}  Nevertheless, the possibility $R_{32}\simeq
R_{33}\simeq 0$ includes other  cases different than the previous
(almost trivial) one.  As mentioned, this represents a small region of
the parameter space, which in principle has no special physical
significance; but for the sake of completeness we will analyze it. $R$
will admit the  parametrization
\bea
\label{R2}
R\simeq\pmatrix{0& \pm \hat c & \hat s \cr   0& \mp \hat s & \hat c \cr \pm 1
&0&0\cr}\;,   \eea
where $\hat \theta$ is an arbitrary complex angle and the sign of the $\pm 1$ entry is not correlated to the signs appearing in the second column. 
Thus, from
eq.(\ref{Ynu}), and keeping only the dominant terms,
\bea
\label{Yaprox}
{\bf Y_\nu}\simeq\pmatrix{0& \pm\sqrt{{\cal M}_1}\hat c  \sqrt{\kappa_2}
&  \sqrt{{\cal M}_1}\hat s \sqrt{\kappa_3}\cr   0& \mp \sqrt{{\cal
M}_2}\hat s  \sqrt{\kappa_2} &  \sqrt{{\cal M}_2}\hat c
\sqrt{\kappa_3}\cr \pm \sqrt{{\cal M}_3 \kappa_1}&0&0\cr}\ U^+\;, \eea
Therefore, the values of $({\bf Y_\nu}^+{\bf Y_\nu})_{ij}$, $i\neq j$,
which are the relevant quantities for $l_i\rightarrow l_j, \gamma$,
depend on the value of $\hat s$. In general, if a unification
condition is imposed for the largest eigenvalue of  ${\bf Y_\nu}^+{\bf
Y_\nu}$, then $({\bf Y_\nu}^+{\bf Y_\nu})_{21}$ is quite large and
BR($\mu\rightarrow e, \gamma$) is above the experimental limit.

Let us illustrate the last point with an example: suppose that $\hat s
= \pm 1$ in eq.(\ref{Yaprox}).
%
%
Then the eigenvalues of ${\bf Y_\nu}^+{\bf Y_\nu}$ are $\{{\cal
M}_3\kappa_1, {\cal M}_2\kappa_2, {\cal M}_1\kappa_3\}$.   If they are
degenerate, ${\bf Y_\nu}^+{\bf Y_\nu}$ is diagonal and we  are in the
above-considered case: all the mixing can be attributed  to the sector
of right-handed neutrinos.  If not,  since $U_{13}\simeq 0$, then
$({\bf Y_\nu}^+{\bf Y_\nu})_{21}\simeq  U_{11}^*U_{21} {\cal
M}_3\kappa_1 + U_{12}^*U_{22} {\cal M}_2\kappa_2$.  This is at least
as large as the matrix element computed in the generic case,
eq.(\ref{Y+Yaprox}), thus leading to unacceptably large
BR($\mu\rightarrow e, \gamma$). For generic values of $\hat s$ the
analysis is more involved, but at the end of the day the conclusion is
similar.

\subsection{$\nu_L$'s hierarchical and $\nu_R$'s degenerate}

As was discussed in subsect.~2.2, if the $R$ matrix is real,  this
scenario is extremely predictive. In particular ${\bf Y_\nu^+}{\bf
Y_\nu}$ is given by eq.(\ref{Ynu+Ynu2}), which  does not depend on the
form of $R$. More precisely
\bea
\label{Ynu+Ynu23}
({\bf Y_\nu^+}{\bf Y_\nu})_{ij}=  {\cal M}\kappa_l
U_{il}U^+_{lj}\simeq   {\cal M}\left[ \kappa_2 U_{i2}U^*_{j2} +
\kappa_3 U_{i3}U^*_{j3} \right] \ ,   \eea
where $\kappa_2\equiv \kappa_{sol}$ and $\kappa_3\equiv \kappa_{atm}$.
The free parameter ${\cal M}$ can be absorbed in the value of the
largest eigenvalue of the ${\bf Y_\nu^+}{\bf Y_\nu}$ matrix,  $|Y_0|^2
\simeq {\cal M}\kappa_3$.  In particular, the matrix element $({\bf
Y_\nu^+}{\bf Y_\nu})_{21}$  becomes
\bea
\label{Ynu+Ynu12hierdeg}
({\bf Y_\nu^+}{\bf Y_\nu})_{21} \simeq    {\cal M}  \kappa_2
U_{22} U_{12}^*  = |Y_0|^2 \frac{\kappa_2}{\kappa_3}U_{22}U_{12}^* \ ,
\eea
where we have used $U_{13}\simeq 0$.  Comparing
(\ref{Ynu+Ynu12hierdeg}) with (\ref{Y+Yaprox}), we see that in this
scenario we expect predictions for  ${\mathrm{BR}}(\mu\rightarrow e,
\gamma)$ and the other processes of the same order as for hierarchical
neutrinos (more precisely, $\sim{\kappa_2/\kappa_3}$ times smaller),
but with no special  textures where the branching ratio becomes
suppressed. All this is  illustrated in Fig.~9. for the LAMSW scenario
using $|Y_0(M_X)|=|Y_t(M_X)|$.

\begin{figure}
\centerline{\vbox{
\psfig{figure=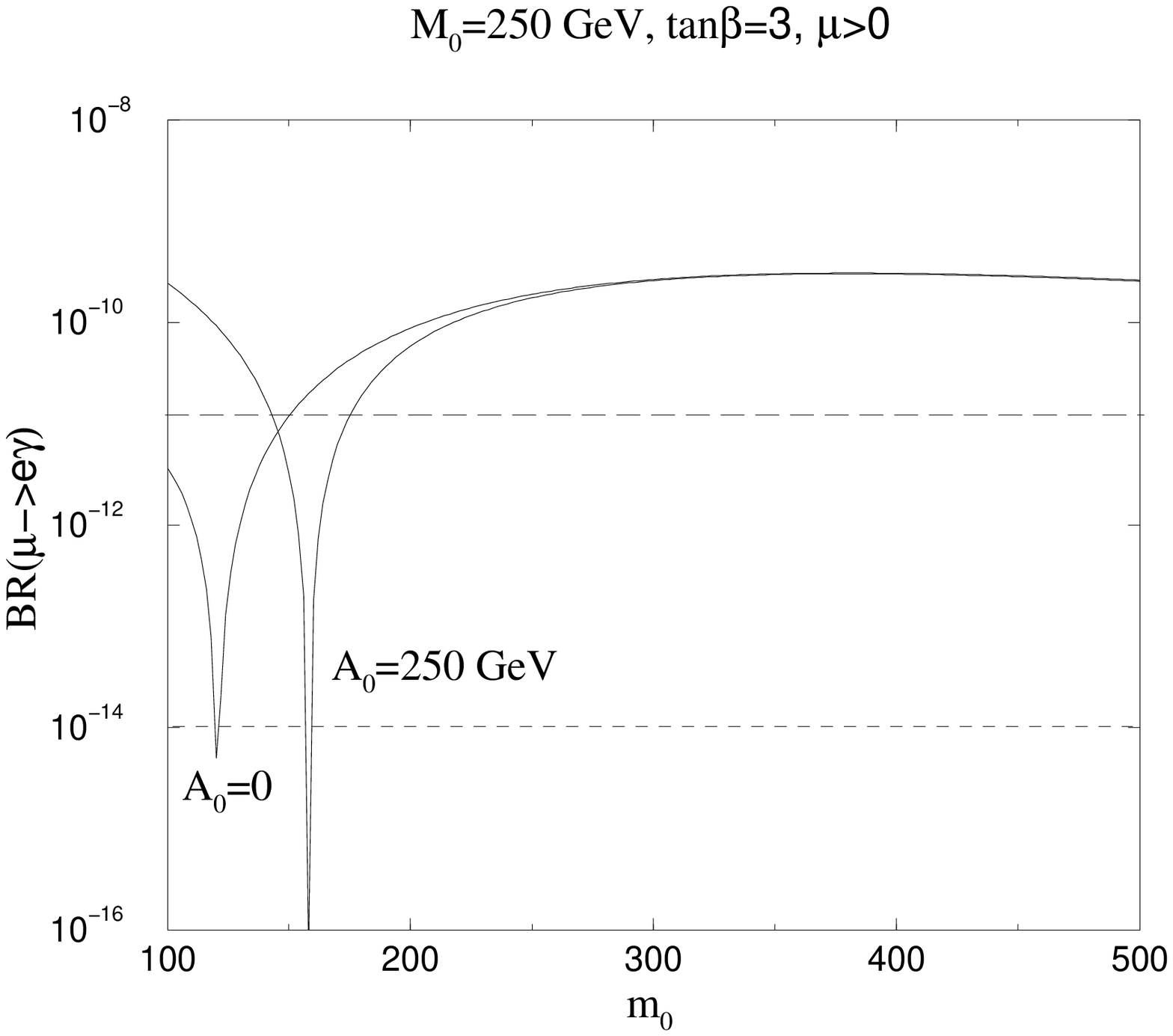,height=10cm}
\psfig{figure=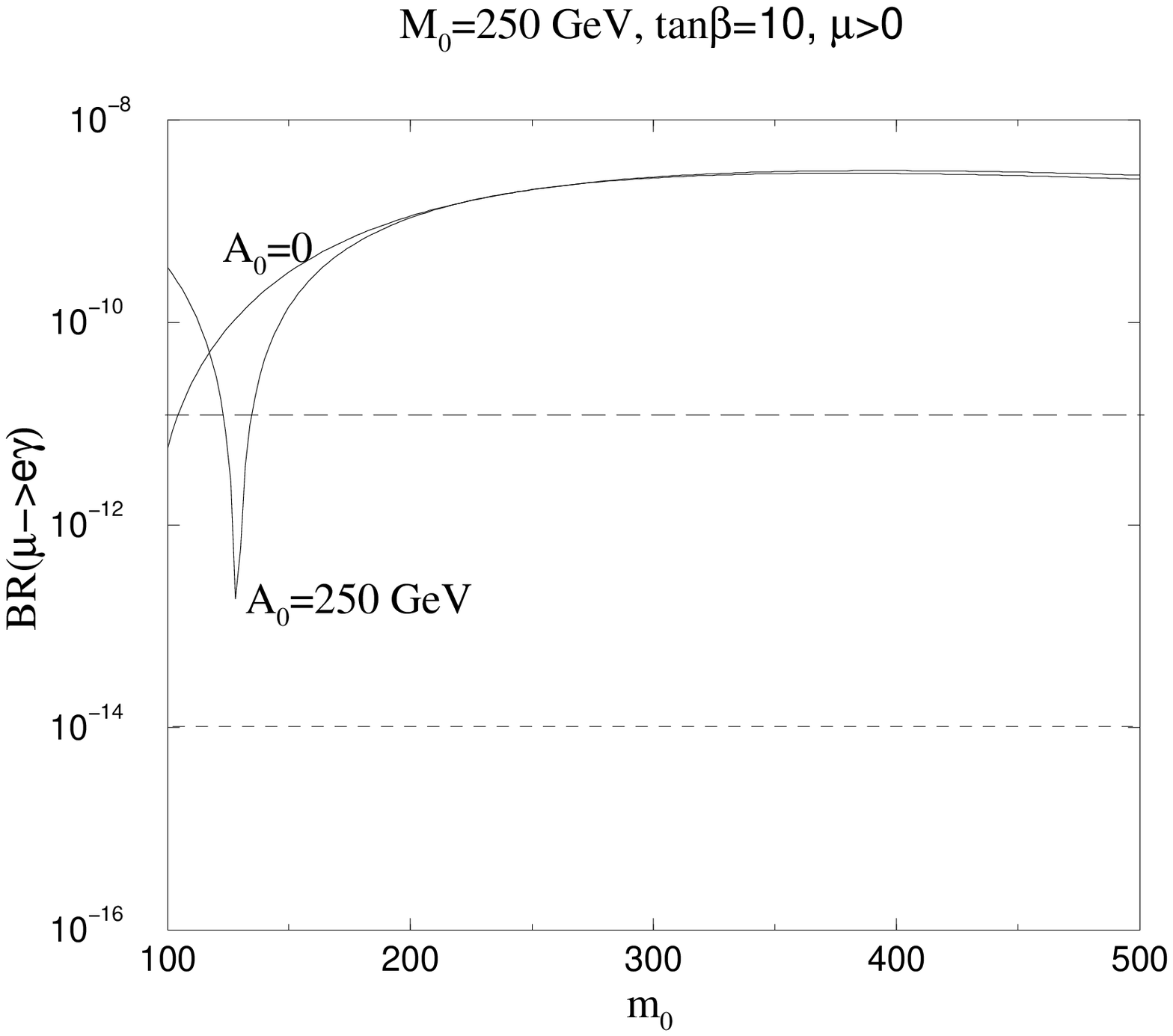,height=10cm}}}
\caption
{\footnotesize
Branching ratio of the process $\mu\rightarrow e, \gamma$
vs. the universal scalar mass, $m_0$, for the case of hierarchical
(degenerate) left (right) neutrinos, $R$ real (see subsect.~4.2) and
two typical sets of 
supersymmetric parameters, as indicated in the plots. 
The dashed lines correspond to the present and forthcoming upper bounds.
A top-neutrino ``unification'' condition has been used to fix
the value of the largest neutrino Yukawa coupling at high energy.
The two plots illustrate two different values of $\tan \beta$.
The curves do not fall below the present bound until  $m_0 > 1.6$ TeV
(uper plot) and  $m_0 > 2.9$ TeV (lower plot).
}
\end{figure}

%
From the plots, it is clear that in this scenario
${\mathrm{BR}}(\mu\rightarrow e, \gamma)$ is already above the
present experimental limits except for a rather small region of 
$m_0$-values which should be probed by the next generation of experiments.

If $R$ is complex, the analysis is more involved since it contains
more arbitrary parameters; but in general the conclusion is the same:
${\mathrm{BR}}(\mu\rightarrow e, \gamma)$ is at least of the same
order as in the real case. To see this note that from the general
equation (\ref{Ynu+Ynu})
\bea
\label{Ynu+YnuRcomplex}
({\bf Y_\nu^+}{\bf Y_\nu})_{21} &=&  {\cal M}\left[U_{2p}
\sqrt{\kappa_p} (R^+R)_{pq}\sqrt{\kappa_q} U^*_{1q}\right]   \nonumber
\\ &\simeq & {\cal M}\left[U_{23} U_{12}^*\sqrt{ \kappa_3 \kappa_2}
R_{q3}^* R_{q2} \ +\   U_{22} U_{12}^* \kappa_2  
R_{q2}^* R_{q2} \right]
\ \ .   \eea
Normally, this is at least of the same order as in the real case,
eq.(\ref{Ynu+Ynu12hierdeg}), as we have checked numerically.  Now
there exists, however, the possibility of a (fine-tuned) cancellation
between the various contributions of (\ref{Ynu+YnuRcomplex}). In
particular such cancellation will occur if all the mixing can be
attributed to the sector of right-handed neutrinos, which implies
${\bf Y_\nu^+}{\bf Y_\nu}$ diagonal.

\subsection{$\nu_L$'s quasi-degenerate}

As it was discussed in subsect.~2.3, in this case it is logical to
assume that ${\cal M}$ has degenerate eigenvalues. If, furthermore,
the $R$ matrix is real, we get from eq.(\ref{Ynu+Ynu})
\bea
\label{Ynu+Ynudeg}
{\bf Y_\nu^+}{\bf Y_\nu}={\cal M} U D_{\kappa} U^+  \ ,  \eea
Incidentally, for $R$ real the requirement that for quasi-degenerate 
neutrinos the two CP phases, $\phi$ and $\phi'$, are opposite in order 
to be consistent with $\nu$-less double $\beta$ decay \cite{Georgi:2000bf},
 implies 
that $U$ and thus  ${\bf Y_\nu}$ is complex, see eqs.(\ref{UV}, 
\ref{Vdef}, \ref{Ynu}).

Since $\kappa_1\simeq\kappa_2\simeq\kappa_3\equiv \kappa$ (typically
$\kappa={\cal O}(1 {\mathrm eV}/v_2^2)$), the eigenvalues of ${\bf
Y_\nu^+}{\bf Y_\nu}$ are quasi-degenerate  $\sim {\cal M} \kappa\equiv
|Y_0|^2$. Clearly, eq.(\ref{Ynu+Ynudeg}) presents a GIM-like
suppression, so the off-diagonal entries of  ${\bf Y_\nu^+}{\bf
Y_\nu}$ are small and proportional to the differences of neutrinos
masses. More precisely, for $i\neq j$
\bea
\label{Ynu+Ynuij}
({\bf Y_\nu^+}{\bf Y_\nu})_{ij}={\cal M} U_{ip}U_{jp}^* \kappa_p =
  {\cal M} U_{ip}U_{jp}^* (\kappa_p-\kappa_3)\ \ .  \eea
In particular, the matrix element $({\bf Y_\nu^+}{\bf Y_\nu})_{21}$,
which is the relevant one for $\mu\rightarrow e, \gamma$, becomes
\bea
\label{Ynu+Ynu12deg}
({\bf Y_\nu^+}{\bf Y_\nu})_{21}={\cal M}  \left[U_{21}U_{11}^*
(\kappa_1-\kappa_3) + U_{22} U_{12}^* (\kappa_2-\kappa_3)\right] \ \ .
\eea
Since $U_{13}\simeq 0$, \footnote{Since we are working here with 
quasi-degenerate neutrinos, one may wonder whether the low-energy condition
$U_{13}\simeq 0$ might change at the ${\cal M}$ scale. However, at is was 
shown in ref.\cite{Casas:2000tg}, the $U_{13}$ element is stable under the RG 
running between ${\cal M}$ and low-energy for any phenomenologically 
viable scenario. Therefore, the argument that follows is completely general.}
from unitarity follows that $U_{21}U_{11}^* \simeq -U_{22} U_{12}^*$. So
\bea
\label{Ynu+Ynu12deg2}
({\bf Y_\nu^+}{\bf Y_\nu})_{21}={\cal M}  \left[U_{21}U_{11}^*
(\kappa_1-\kappa_2) \right] = |Y_0|^2 U_{21}^* U_{11} \frac{\Delta
\kappa_{sol}^2}{\kappa^2} \ \ .  \eea
The factor $\frac{\Delta \kappa_{sol}^2}{\kappa^2}$ is at most $\sim
10^{-5}$ (this occurs for the LAMSW  solution to the solar neutrino
problem), which implies a drastic suppression of $({\bf Y_\nu^+}{\bf
Y_\nu})_{21}$, and thus of BR($\mu\rightarrow e, \gamma$) in this kind
of scenario. More quantitatively, eq.(\ref{Ynu+Ynu12deg2}) is
identical to  eq.(\ref{Ynu+Ynu12hierdeg}) (i.e. the scenario where
$\nu_L$'s are hierarchical and $\nu_R$'s degenerate), multiplied by
$\kappa^{-2}\sqrt{\Delta \kappa_{sol}^2 \Delta \kappa_{atm}^2}$.  This
is a factor $\sim 10^{-4}$ for the LAMSW, which means that all the
plots representing BR($\mu\rightarrow e, \gamma$) in that scenario
(Fig.~9) are identical here, but with the vertical axis re-scaled eight
orders of magnitude smaller.

If the $R$ matrix is complex, then $R^+R$ is in general non-diagonal,
so the ${\bf Y_\nu^+}{\bf Y_\nu}$ matrix becomes
\bea
\label{Ynu+Ynudeg2}
{\bf Y_\nu^+}{\bf Y_\nu}={\cal M} U D_{\sqrt{\kappa}}R^+R
D_{\sqrt{\kappa}} U^+  \ ,  \eea
which may have sizeable off-diagonal entries. Hence, even for
quasi-degenerate neutrinos, $({\bf Y_\nu^+}{\bf Y_\nu})_{21}$, and
thus BR($\mu\rightarrow e, \gamma$), could be very large if $R$ is
complex.

\vspace{0.3cm}
\noindent
Finally, we would like to comment on the scenario where the neutrinos
are partially degenerate, i.e. when their masses present an inverse
hierarchy: $\kappa_3\ll \kappa_1\simeq\kappa_2\equiv \kappa$.  [It is
convenient to maintain $\kappa_3$ as the most split mass eigenvalue,
so that the interpretation of $\theta_{23}$, $\theta_{12}$ as the
atmospheric and solar angles respectively remains valid.]  Notice  that
in this case $\kappa^2\simeq \Delta \kappa_{atm}^2$.

Now, the discussion of the size of $({\bf Y_\nu^+}{\bf Y_\nu})_{21}$, and
thus of BR($\mu\rightarrow e, \gamma$),
is similar to that in the complete degenerate case, though the 
conclusions are different. In particular, suppose that ${\cal M}$ has 
degenerate eigenvalues, 
$D_{\cal M}={\mathrm diag}({\cal M},{\cal M},{\cal M})$, and that
the $R$ matrix is real. Then eq.(\ref{Ynu+Ynudeg}) holds, so 
${\bf Y_\nu^+}{\bf Y_\nu}$ has two large quasi-degenerate eigenvalues
$\sim {\cal M} \kappa\equiv |Y_0|^2$. Likewise, all the discussion 
between the eqs. (\ref{Ynu+Ynuij})--(\ref{Ynu+Ynu12deg2}) 
and the equations themselves remain valid. Therefore, the matrix 
element $({\bf Y_\nu^+}{\bf Y_\nu})_{21}$ becomes
\bea
\label{Ynu+Ynu12deg3}
({\bf Y_\nu^+}{\bf Y_\nu})_{21}=
 |Y_0|^2U_{21}U_{11}^*\frac{\Delta \kappa_{sol}^2}{\kappa^2}
=|Y_0|^2U_{21}U_{11}^*\frac{\Delta \kappa_{sol}^2}{\Delta \kappa_{atm}^2}
\ \ .
\eea
Again, this equation is identical to  eq.(\ref{Ynu+Ynu12hierdeg}),
multiplied now by 
$\sqrt{\frac{\Delta \kappa_{sol}^2}{ \Delta \kappa_{atm}^2}}$. 
This suppression factor is not as strong as before. Namely, it is 
$\sim 10^{-1}$ for the LAMSW, which means that 
all the plots representing BR($\mu\rightarrow e, \gamma$) in
this scenario are as in the one where 
$\nu_L$'s are hierarchical and $\nu_R$'s degenerate, re-scaled by a 
factor $\sim 10^{-2}$. Hence, from Fig.~9 it is clear
that BR($\mu\rightarrow e, \gamma$) for the partially degenerate case
should be testable within the next generation of experiments.

The matrix element $({\bf Y_\nu^+}{\bf Y_\nu})_{31}$ gets a 
similar suppression. However $({\bf Y_\nu^+}{\bf Y_\nu})_{32}$ 
is not suppressed at all:
\bea
\label{Ynu+Ynu32deg}
({\bf Y_\nu^+}{\bf Y_\nu})_{32}={\cal M} 
\left[U_{31}U_{21}^* (\kappa_1-\kappa_3) +
U_{32} U_{22}^*(\kappa_2-\kappa_3)\right]
\simeq {\cal M}\kappa\left[U_{31} U_{21}^*+U_{32} U_{22}^*\right]
\ \ ,
\eea
where ${\cal M}\kappa = |Y_0|^2$. This is a similar size to that from
the hierarchical scenarios. So, for the
partially degenerate case,
BR($\tau\rightarrow \mu, \gamma$)
should also be testable in the next generation of experiments.

Again, if the $R$ matrix is complex ${\bf Y_\nu^+}{\bf Y_\nu}$ 
is given by eq.(\ref{Ynu+Ynudeg2}) and we expect much larger values for 
$({\bf Y_\nu^+}{\bf Y_\nu})_{21}$, $({\bf Y_\nu^+}{\bf Y_\nu})_{31}$.

\section{Summary of predictions for BR($l_i\rightarrow l_j, \gamma$)}

If the origin of the neutrino masses is a see-saw mechanism, 
implemented in a supersymmetric theory, the soft breaking terms 
get off-diagonal contributions through the RG running, even 
if one starts with a universality condition for the soft terms.
E.g. for the slepton doublet mass-matrix, at the leading-log approximation,
\bea
\label{summary1} 
\left(m_L^2\right)_{ij} & \simeq & \frac{-1}{8\pi^2}(3m_0^2 + A_0^2)
({\bf Y_\nu^+}{\bf Y_\nu})_{ij} \log\frac{M_X}{M}\ ,
\eea
where  ${\bf Y_\nu}$ is the matrix of neutrino Yukawa couplings, $M$ is the typical (Majorana) mass of the right-handed neutrinos and
$M_X$ is the initial scale at which universality is imposed. Therefore, the predictions on  BR($l_i\rightarrow l_j, \gamma$) are directly linked to 
the size of $({\bf Y_\nu^+}{\bf Y_\nu})_{ij}$. 
Moreover, the larger (smaller) the initial scale at which 
universality is imposed the larger the branching ratios.
Also, due to the structure of the diagrams,  BR($l_i\rightarrow l_j, \gamma$) has a strong dependence on $\tan \beta$ ($\sim \tan^2 \beta$).
The most general form of
${\bf Y_\nu}$ and thus of ${\bf Y_\nu^+}{\bf Y_\nu}$, in terms of measurable low-energy parameters has been determined in sect.~2. In particular, working in the flavour basis in which the gauge
interactions and the charged-lepton Yukawa matrix, $\bf{Y_e}$,
are flavour-diagonal,
\bea
\label{summary2}
{\bf Y_\nu^+}{\bf Y_\nu}= U D_{\sqrt{\kappa}} R^+ D_{\cal M} R
D_{\sqrt{\kappa}} U^+  
\eea
Here $D_{\sqrt{\kappa}}$ is the diagonal matrix of positive neutrino mass 
eigenvalues ($\kappa = {\cal M}_\nu/ \langle H_2^0\rangle^2$), 
while $U$ contains the  mixing angles and CP phases 
[see eqs.(\ref{Udiag}, \ref{CKM})]. In addition, 
${\bf Y_\nu^+}{\bf Y_\nu}$ depends on unknown parameters: the 
 three  positive mass
eigenvalues of the righthanded neutrinos (contained in $D_{\cal M}$) and
the three complex parameters
defining the complex orthogonal matrix $R$.
In practical cases, however,
the number  of relevant free parameters becomes drastically reduced.

Next, we enumerate 
the possible low-energy scenarios and the corresponding predictions for 
$({\bf Y_\nu^+}{\bf Y_\nu})_{ij}$ and BR($l_i\rightarrow l_j, \gamma$), 
as has been worked out in detail in sects. 2, 3, 4.

\subsection*{$\nu_L$'s and $\nu_R$'s completely hierarchical}    

\begin{itemize}

\item 
If $R$ is a generic matrix, $({\bf Y_\nu^+}{\bf Y_\nu})_{ij}$
is given by eq.(\ref{Y+Yjer}). Using the parametrization of $R$ given 
in eq.(\ref{R}), which is sufficiently general for this case, one
obtains in particular
\bea
\label{summary3} 
({\bf Y_\nu^+}{\bf Y_\nu})_{21}\sim \frac{|{Y}_0|^2}{|\hat
s_1|^2\kappa_2+ |\hat c_1|^2\kappa_3} \hat c_1^*\hat s_1 
\sqrt{\kappa_3\kappa_2}U_{23}  U_{12}^* 
\eea
Here $|{Y}_0|^2$ is the largest eigenvalue of ${\bf Y_\nu^+}{\bf
Y_\nu}$ [see eq.(\ref{unifcond})] and $\hat \theta_1$ is an 
arbitrary complex angle.

For the LAMSW scenario the previous equation generically gives
BR($\mu\rightarrow e, \gamma$) above the {\em present} experimental
limits,   at least for ${Y}_0 = {\cal O}(1)$, as it occurs in the
unified scenarios.  This is illustrated in Figs. 3, 4, 5, 6
using $|Y_0(M_X)|=|Y_t(M_X)|$, with $M_X=2\times
10^{16}$ GeV.  But even for
${Y}_0 = {\cal O}(10^{-1})$, most of the parameter space  will be
probed in the forthcoming generation of experiments experiments (see
Fig.~7).

\item The only exceptions to the previous result are

  \begin{itemize}

  \item If $\hat \theta_1$ takes one of the two values given in
       eqs.(\ref{puntos1}, \ref{puntos2}), which correspond to the
       textures of eqs.(\ref{textures1}, \ref{textures2}).  Then,
       BR($\mu\rightarrow e, \gamma$) is drastically reduced, but
       other processes, as ${\mathrm{BR}}(\tau\rightarrow \mu,
       \gamma)$, are not, normally lying above the forthcoming experimental
       upper bound, see Fig.~8.

 \item If $R$ is such that ${\bf Y_\nu^+}{\bf Y_\nu}$ is
       diagonal. This means that in a certain basis $\bf{Y_e}$ and
       ${\bf Y_\nu}$ are diagonal whereas  the right-handed Majorana
       mass matrix, ${\cal M}$, is not. This requires a very special
       form of $R$, which in particular has $R_{32},R_{33} \simeq 0$.
       Of course, in this special case there is no RG  generation of
       non-diagonal soft terms, and the predictions for
       BR($l_i\rightarrow l_j, \gamma$) are as in the SM,
       i.e. negligible.

  \end{itemize}

\end{itemize}

\subsection*{$\nu_L$'s hierarchical and $\nu_R$'s degenerate}

\begin{itemize}

\item  If $R$ is real, this scenario is very predictive. 
 $({\bf Y_\nu^+}{\bf Y_\nu})_{ij}$ is given by eq.(\ref{Ynu+Ynu23}).
In particular, taking into account $U_{13}\simeq 0$,
\bea
\label{summary4}
({\bf Y_\nu^+}{\bf Y_\nu})_{21} \simeq   
{\cal M}  \kappa_2 U_{22}  U_{12}^* 
= |Y_0|^2 \frac{\kappa_2}{\kappa_3}U_{22} U_{12}^*\ ,
\eea
where $|Y_0|^2 \simeq {\cal M}\kappa_3$ is the largest eigenvalue 
of ${\bf Y_\nu^+}{\bf Y_\nu}$.

The corresponding ${\mathrm{BR}}(\mu\rightarrow e, \gamma)$ is shown
in  Fig.~9. for the LAMSW scenario, and using $|Y_0(M_X)|=|Y_t(M_X)|$.
\footnote{For other values of $|Y_0|$ note that 
${\mathrm{BR}}(\mu\rightarrow e, \gamma)$) $\sim$ $|Y_0|^4$.
Actually, this dependence is milder, since the larger $|Y_0(M_X)|$, 
the faster it decreases with the scale.} 
The branching ratio turns out to be  already above the 
present experimental limits except for a rather small region of $m_0$ 
values which should be probed by the next generation of experiments.
Here are not special textures where the branching ratio becomes suppressed.

\item  
If $R$ is complex, the analysis is more involved since it contains
more arbitrary parameters; $({\bf Y_\nu^+}{\bf Y_\nu})_{21}$ is given by
eq.(\ref{Ynu+YnuRcomplex}). But 
in general the conclusion is the same:
${\mathrm{BR}}(\mu\rightarrow e, \gamma)$ is at least of the same
order as in the real case. Now there exists, however, the possibility of a
(fine-tuned) cancellation between the various contributions of
(\ref{Ynu+YnuRcomplex}). In particular such cancellation will occur if
all the mixing can be attributed to the sector of right-handed
neutrinos, which implies ${\bf Y_\nu^+}{\bf Y_\nu}$ diagonal.

\end{itemize}

\subsection*{$\nu_L$'s quasi-degenerate}

\begin{itemize}

\item  If $R$ is real, $({\bf Y_\nu^+}{\bf Y_\nu})_{ij}$ is given 
by eq.(\ref{Ynu+Ynudeg}). In particular $({\bf Y_\nu^+}{\bf Y_\nu})_{21}$
is given by eq.(\ref{Ynu+Ynu12deg2})
\bea
\label{summary5}
({\bf Y_\nu^+}{\bf Y_\nu})_{21}={\cal M} 
\left[U_{21} U_{11}^*(\kappa_1-\kappa_2) \right]
= |Y_0|^2 U_{21}U_{11}^* \frac{\Delta \kappa_{sol}^2}{\kappa^2}
\ \ .
\eea
where, again, $|Y_0|^2 \simeq {\cal M}\kappa$ is the largest eigenvalue 
of ${\bf Y_\nu^+}{\bf Y_\nu}$. This equation is identical to
eq.(\ref{summary4}), multiplied by 
$\kappa^{-2}\sqrt{\Delta \kappa_{sol}^2 \Delta \kappa_{atm}^2}$. 
This is a factor $\sim 10^{-4}$ for the LAMSW.
Therefore all the plots representing BR($\mu\rightarrow e, \gamma$) 
in the previous scenario (Fig.~9)
are valid here, but with the vertical axis re-scaled eight 
orders of magnitude smaller. Consequently,  BR($\mu\rightarrow e, \gamma$)
is naturally suppressed below the present (and even forthcoming) limits.

\item 
If $R$ is complex, ${\bf Y_\nu^+}{\bf Y_\nu}$ is given  by
eq.(\ref{Ynu+Ynudeg2}), which may have sizeable off-diagonal
entries. Hence,  BR($\mu\rightarrow e, \gamma$),
could be very large in this case.

\item
If the (quasi-) degeneracy is only partial: 
$\kappa_3\ll \kappa_1\simeq\kappa_2\equiv \kappa\sim 
\sqrt{\Delta \kappa_{atm}^2}$, 
$({\bf Y_\nu^+}{\bf Y_\nu})_{21}$
is given (for $R$ real) by eq.(\ref{Ynu+Ynu12deg3}). 
Again, this equation is identical to  
eq.(\ref{summary4}), multiplied now by 
$\sqrt{\frac{\Delta \kappa_{sol}^2}{ \Delta \kappa_{atm}^2}}$. 
This represents a suppression factor 
$\sim 10^{-1}$ for the LAMSW, which means that 
all the plots representing BR($\mu\rightarrow e, \gamma$) 
are as those of  Fig.~9, re-scaled by a 
factor $\sim 10^{-2}$. As a consequence, 
BR($\mu\rightarrow e, \gamma$) for this partially degenerate scenario
should be testable within the next generation of experiments.
The conclusion is similar for BR($\tau\rightarrow \mu, \gamma$).

For generic complex $R$, the value of BR($\mu\rightarrow e, \gamma$)
does not get any suppression and falls naturally above the present
experimental limits.

\end{itemize}

\section{Concluding remarks}

If the origin of the neutrino masses is a supersymmetric see-saw, which 
is probably the most attractive scenario to explain their smallness,
then the leptonic soft breaking terms acquire off-diagonal contributions 
through the RG running, which drive non-vanishing 
BR($l_i\rightarrow l_j, \gamma$). These contributions
are proportional to $({\bf Y_\nu^+}{\bf Y_\nu})_{ij}$, where 
${\bf Y_\nu}$ is the neutrino Yukawa matrix,

Therefore, in order to make predictions for these branching ratios, one has
first to determine the most general form of 
${\bf Y_\nu}$ and ${\bf Y_\nu^+}{\bf Y_\nu}$, compatible with all 
the phenomenological requirements. 
This has been the first task of this paper, and the result is summarized in 
eqs.(\ref{Ynu}, \ref{Ynu+Ynu}). 

Then, we have shown that the predictions for BR($\mu\rightarrow e, \gamma$) 
are normally {\em above} the {\em present} experimental limits if the three 
following conditions occur

\begin{enumerate}

\item The solution to the solar neutrino problem is the LAMSW, 
as favoured by the most recent analyses.

\item  ${Y}_0(M_X) = {\cal O}(1)$, where $|Y_0|^2$ is the largest
eigenvalue of $({\bf Y_\nu^+}{\bf Y_\nu})$. This occurs e.g. in most 
grand-unified scenarios.

\item The soft-breaking terms are generated at a high-energy scale, e.g.
$M_X$, above the Majorana mass of the right-handed neutrinos, $M$.

\end{enumerate}

\noindent
These conditions are very plausible. In our opinion, the most natural
scenarios fulfill them, but certainly there exists other
possibilities. E.g. concerning condition 1, the solution to the solar 
neutrino problem could
be LOW. In this case,  BR($\mu\rightarrow e, \gamma$) is ($\sim 20$ times) 
smaller, but
still testable in the forthcoming experiments. Condition 2 is
satisfied in $SO(10)$ models, but it could be unfulfilled in $SU(5)$
models or in string scenarios with no GUT group.  Still, we have noted
in this work that even for  ${Y}_0 = {\cal O}(10^{-1})$, most of the
parameter space  will be probed in the forthcoming generation of
experiments (year 2003).  Finally, condition 3 seems a
natural one, as the see-saw mechanism requires the existence of
high-energy scales, and it is hardly compatible e.g. with models where
the fundamental scale is ${\cal O}$ (TeV). 
Actually, if the initial scale at which universality is imposed is larger than the ordinary GUT $M_X$ scale (as it is common in gravity mediated supersymmetry breaking), the predictions for the branching ratios increase.
However, it may happen that
supersymmetry is broken at a scale below $M$. This is the case of
gauge-mediated scenarios, where there would be no generation of
off-diagonal leptonic soft terms through the RG running. In this
sense, the gauge-mediated models are free from $\mu\rightarrow e,
\gamma$ constraints and represent a way-out to the previous conclusion.

\vspace{0.3cm}
\noindent
Even under the previous 1--3 conditions, there are physical scenarios
compatible with the present  BR($\mu\rightarrow e, \gamma$) 
experimental limits. Namely

\begin{itemize}

\item Whenever all the leptonic flavour violation can be attributed to
the sector of right-handed neutrinos. This happens when there exists a
basis in which the gauge couplings, $\bf{Y_e}$ and ${\bf Y_\nu}$ get  
simultaneously flavour-diagonal, while the right-handed Majorana matrix, 
${\cal M}$, gets
non-diagonal. In this case there is no RG generation of non-diagonal
soft terms, and the predictions for BR($l_i\rightarrow l_i, \gamma$)
are as in the SM, i.e. negligible.

This possibility certainly exists and might be considered an
attractive one.  However, it should be noticed that it does {\em not} occur
in the quark sector since the $\bf{Y_u}$ and ${\bf Y_d}$ Yukawa
matrices cannot be  simultaneously diagonalized in a basis where  the
gauge couplings remain diagonal. In our opinion, it would be a 
big chance that this happens in the leptonic sector.

\item In the scenario where  both, the left-handed and the
right-handed neutrinos have hierarchical masses, there are two special
textures, shown in eqs.(\ref{textures1}, \ref{textures2}), 
where BR($\mu\rightarrow e,
\gamma$) becomes  suppressed. The first one is probably the most
appealing,  since it takes place for the simple case $R={\bf 1}$ in
eq.(\ref{Ynu}).  This is equivalent to assume that there exists a basis
of $L_i$  and ${\nu_R}_i$ in which ${\bf Y_\nu}$ and ${\cal M}$ are
simultaneously diagonal, though not ${\bf Y_e}$. In other words, all
the leptonic flavour violation can be attributed to the sector of
charged leptons.

Then, we have shown that for this privileged texture
BR($\tau\rightarrow \mu, \gamma$) is not suppressed, lying in general above the
forthcoming experimental upper bound.

\item If the left-handed neutrinos are quasi-degenerate, then
BR($l_i\rightarrow l_j, \gamma$) are naturally suppressed. Namely, if
the $R$ matrix in eq.(\ref{Ynu}) is real, there is a GIM-like
mechanism that  drives BR($l_i\rightarrow l_j, \gamma$) below the
experimental limits, even the forthcoming ones. (This is not the
case, however, if $R$ is complex.)

If the degeneracy is partial (case of inverse hierarchy of neutrino
masses), BR($\mu\rightarrow e, \gamma$) is somewhat suppressed, but
still testable in the forthcoming experiments. The latter is also true
for BR($\tau\rightarrow \mu, \gamma$)

\end{itemize}

In our opinion, if the above 1--3 conditions are fulfilled, the scenario 
of quasi-degenerate neutrinos and the one with gauge mediated supersymmetry 
breaking  represent the most plausible explanations
to the absence of $\mu\rightarrow e, \gamma$ observations, specially if 
the absence persists after the next generation of experiments.

As a final conclusion, the discovery of neutrino oscillations makes
much more plausible the possibility of observing
lepton-flavour-violation processes, specially $\mu\rightarrow e,
\gamma$, if the theory is supersymmetric and  the neutrino masses are
generated by a see-saw mechanism.  Large regions of the parameter
space are {\em already excluded} on these grounds, and there exists great
chances to observe $\mu\rightarrow e, \gamma$ in the near future (PSI,
2003). This means that, hopefully, we will have signals of
supersymmetry before LHC.

\section*{Acknowledgements}

The work of J.A.C. was supported in part by the CICYT (contract
AEN98-0816).
A.I. would like to thank JoAnne Hewett and Gudrun Hiller, 
and specially Sacha Davidson and Graham Ross for very interesting 
discussions and invaluable suggestions. A.I. would also like to thank 
the SLAC theory group for hospitality during the first stages 
of this work.

\section*{Appendix}

In the appendix, we summarize the relevant formulas concerning
RGEs and the coefficients of the 
amplitude appearing in eq.(\ref{T}).

Between $M_X$ and $M$ the evolution of the Yukawa couplings is given by
\bea
\label{rg2}
\frac{d {\bf Y_\nu}}{dt}= -\frac{1}{16\pi^2} {\bf Y_\nu}\left[\left(
3 g_2^2+\frac{3}{5}g_1^2-{\mathrm T_2}
\right){\bf I_3}-\left( 3 {\bf
Y_{\nu}^+}  {\bf Y_\nu}+{\bf Y_e^+ Y_e}\right) \right], \eea
\bea
\label{rg3}
\frac{d {\bf Y_e}}{dt}= -\frac{1}{16\pi^2} {\bf Y_e} \left[\left(
3 g_2^2+\frac{9}{5}g_1^2-{\mathrm T_1}
\right){\bf I_3}-\left(
{\bf Y_\nu^+} {\bf Y_\nu}+ 3  {\bf Y_e^+ Y_e}\right) \right],
\eea
\bea 
\frac{d {\bf Y_U}}{dt}= -\frac{1}{16\pi^2}{\bf Y_U} \left[\left(
\frac{13}{15}g_1^2+3 g_2^2+ \frac{16}{3} g_3^2- T_2
\right){\bf I}_3
- 3 {\bf Y_U^+ Y_U}- {\bf Y_D^+ Y_D} \right],
\label{YtRGE} 
\eea
where
\bea
{\mathrm T_1}={\mathrm Tr}(
3 {\bf Y_D^+ Y_D}+{\bf Y_e^+ Y_e}),\;\;\;
{\mathrm T_2}={\mathrm Tr}(
3{\bf Y_U^+ Y_U}+{\bf Y_\nu^+ Y_\nu}),
\eea
and
\bea
\label{rg4}
\frac{d {\cal M}}{dt}=\frac{1}{8\pi^2}\left[{\cal M} ({\bf Y_\nu}
{\bf Y_\nu}^+)^T+{\bf Y_\nu} {\bf Y_\nu}^+ {\cal M}\right];
\eea 
The renormalization group equations for the soft masses and 
trilinear terms are:
\bea
\frac{d {\bf m^2_L} }{d t} &=&
\frac{1}{16 \pi^2} \left [
\left({\bf m^2_L} {\bf Y_e^{+}} {\bf Y_e}
+{\bf Y_e^{+}} {\bf Y_e} {\bf m^2_L} \right)
+\left({\bf m^2_L} {\bf Y_{\nu}^+ Y_{\nu}}
+{\bf Y_{\nu}^+ Y_{\nu}} {\bf m^2_L} \right)
\right.
\nonumber \\
&&+2 \left( {\bf Y_e^{+}} {\bf m^2_e} {\bf Y_e}
+{ m}^2_{H_1}{\bf Y_e^{+}}{\bf Y_e}
+{\bf A_e^+} {\bf A_e} \right)
+2 \left( {\bf Y_{\nu}^{+}} {\bf m^2_{\nu}} {\bf Y_{\nu}}
+{ m}^2_{H_2}{\bf Y_{\nu}^{+}} {\bf Y_{\nu}}
+{\bf A_{\nu}^{+}} {\bf A_{\nu}} \right)
\nonumber \\
&&
\left.
-\left (\frac{6}{5}g_1^2 \left| M_1 \right|^2
+6 g_2^2 \left| M_2 \right|^2 \right) {\bf I_3}
-\frac{3}{5} g_1^2 S {\bf I_3}
\right ],
\label{rgmL} \\
\frac{d {\bf m^2_e} }{d t}&=&
\frac{1}{16 \pi^2} \left [
2 \left({\bf m^2_e} {\bf Y_e} {\bf Y_e^{+}}
+{\bf Y_e} {\bf Y_e^{+}} {\bf m^2_e} \right)
+4 \left( {\bf Y_e} {\bf m^2_L} {\bf Y_e^{+}}
+{ m}^2_{H_1}{\bf Y_e} {\bf Y_e^{+}}
+{\bf A_e} {\bf A_e^{+}}\right)
\right.
\nonumber \\
&&\left.
-\frac{24}{5} g_1^2 \left| M_1 \right|^2 {\bf I_3}
+\frac{6}{5}g_1^2 S {\bf I_3}
\right],
\\
\frac{d {\bf m^2_{\nu}} }{d t} &=&
\frac{1}{16 \pi^2} \left [
2 \left({\bf m^2_{\nu}} {\bf Y_{\nu}} {\bf Y_{\nu}^+}
+{\bf Y_{\nu}}{\bf Y_{\nu}^{+}} {\bf m^2_{\nu}} \right)
+4 \left( {\bf Y_{\nu}} {\bf m^2_L} {\bf Y_{\nu}^+}
+{ m}^2_{H_2}{\bf Y_{\nu}} {\bf Y_{\nu}^{+}}
+{\bf A_{\nu}} {\bf A_{\nu}^+}\right)
\right],
\\
\frac{d m^2_{H_2} }{d t} &=&
\frac{1}{16 \pi^2} \left [6{\mathrm Tr}\left({\bf Y_U^{+}
 ( m_Q^2+m_U^2}+ m^2_{H_2}{\bf I_3}) {\bf Y_U + A_U^{+} A_U}\right) \right.
\nonumber \\
&&
\left.
+2 {\mathrm Tr} \left( {\bf Y_{\nu}^{+} (m_L^2+m^2_{\nu}}
+ m^2_{H_2}{\bf I_3}){\bf Y_{\nu}+ A_{\nu}^{+} A_{\nu}}\right) \right.
\nonumber \\
&&
\left.
-\left (\frac{6}{5}g_1^2 \left| M_1 \right|^2
+6 g_2^2 \left| M_2 \right|^2 \right)
+\frac{3}{5} g_1^2 S
\right],
\\
\frac{d {\bf A_e} }{d t}&=&
\frac{1}{16 \pi^2} \left[ \left\{
-\frac{9}{5} g_1^2 -3 g_2^2+ 3 {\mathrm Tr}({\bf Y_D^+} {\bf Y_D})
+{\mathrm Tr}({\bf Y_e^+} {\bf Y_e}) \right \} {\bf A_e}
\right.
\nonumber \\
&&+2 \left\{
-\frac{9}{5} g_1^2 M_1 -3 g_2^2 M_2 + 3 {\mathrm Tr}({\bf Y_D^+} {\bf A_D})
+{\mathrm Tr}({\bf Y_e^+} {\bf A_e}) \right \} {\bf Y_e}
\nonumber \\
&&\left.
+4 ({\bf Y_e} {\bf Y_e^+} {\bf A_e}) + 5 ({\bf A_e} {\bf Y_e^+} {\bf Y_e})
+2({\bf Y_e} {\bf Y_{\nu}^+} {\bf A_{\nu}}) +
 ({\bf A_e} {\bf Y_{\nu}^+} {\bf Y_{\nu}})
\right],
\label{rgAe}
\\
\frac{d {\bf A_{\nu}}}{d t} &=&
\frac{1}{16 \pi^2} \left[ \left\{
-\frac{3}{5}g_1^2 -3g_2^2 +3 {\mathrm Tr}({\bf Y_U^+} {\bf Y_U})
+{\mathrm Tr}({\bf Y_{\nu}^+} {\bf Y_{\nu}}) \right \} {\bf A_{\nu}}
\right.
\nonumber \\
&&+2 \left\{
-\frac{3}{5} g_1^2 M_1 -3 g_2^2 M_2 + 3 {\mathrm Tr}({\bf Y_U^+} {\bf A_U})
+{\mathrm Tr}({\bf Y_{\nu}^+} {\bf A_{\nu}}) \right \} {\bf Y_{\nu}}
\nonumber \\
&&\left.
+4({\bf Y_{\nu}} {\bf Y_{\nu}^+}{\bf A_{\nu}})
+5({\bf A_{\nu}}{\bf Y_{\nu}^+} {\bf Y_{\nu}})
+2({\bf Y_{\nu}} {\bf Y_e^+}{\bf A_e})
+({\bf A_{\nu}} {\bf Y_e^+} {\bf Y_e})
\right],
\\
\frac{d {\bf A_U}}{d t} &=&
\frac{1}{16 \pi^2} \left[ \left\{
-\frac{13}{15}g_1^2 -3g_2^2 -\frac{16}{3}g_3^2 +
3 {\mathrm Tr}({\bf Y_U^+} {\bf Y_U})
+{\mathrm Tr}({\bf Y_{\nu}^+} {\bf Y_{\nu}}) \right \} {\bf A_U}
\right.
\nonumber \\
&&
+2 \left\{
-\frac{13}{15} g_1^2 M_1 -3 g_2^2 M_2 -\frac{16}{3} g_3^2 M_3 
+ 3 {\mathrm Tr}({\bf Y_U^+} {\bf A_U})
+{\mathrm Tr}({\bf Y_{\nu}^+} {\bf A_{\nu}}) \right \} {\bf Y_U}
\nonumber \\
&&\left.
+4({\bf Y_U} {\bf Y_U^+}{\bf A_U})
+5({\bf A_U}{\bf Y_U^+} {\bf Y_U})
+2({\bf Y_U} {\bf Y_D^+}{\bf A_D})
+({\bf A_U} {\bf Y_D^+} {\bf Y_D})
\right],
\end{eqnarray}
where
\begin{eqnarray}
S={\mathrm Tr}({\bf m_{ Q}^2} +{\bf m_{ d}^2} -2{\bf m_{ u}^2}
-{\bf m_{ L}^2} +{\bf m_{ e}^2)}
- { m}^2_{H_1}+{ m}^2_{H_2}.
\eea
Here $g_2$ and $g_1$ are the $SU(2)_L$ and $U(1)_Y$ gauge coupling 
constants, and ${\bf Y_{U,D,e,\nu}}$ are the Yukawa matrices for up quarks, 
down quarks, charged leptons and neutrinos respectively. The RGEs for 
${\bf Y_D}$ and the rest 
of the soft masses or trilinear terms are the same as in the MSSM.
Below $M$, the RGEs are the same, except that the right neutrinos decouple, 
and hence ${\bf Y_{\nu}}$ disappears from the equations.

\vspace{0.3cm}
On the other hand, following ref.\cite{Hisano:1996cp}
the amplitude for the process $l_i\rightarrow l_j \gamma$ 
can be written as
\begin{eqnarray}
T= \epsilon^{\alpha} {\bar l_j} 
m_{l_i} i \sigma_{\alpha \beta} q^\beta (A_L P_L + A_R P_R)
l_i,
\label{Penguin}
\end{eqnarray}
\begin{eqnarray}
A_{L, R}&=&A^{(c)}_{L, R}+A^{(n)}_{L, R}
\nonumber
\end{eqnarray}

\bea
A^{(c)}_L&=&-\frac{e}{16 \pi^2}\frac{1}{m^2_{\tilde{\nu}_X}}
\left[
 C_{jAX}^{L} C_{iAX}^{L*} I_1(M^2_{\tilde{\chi}^-_A}/
m^2_ {\tilde{\nu}_X})+C_{jAX}^{L} C_{iAX}^{R*} 
\frac{M_{\tilde{\chi}_A^-}}{m_{l_i}} I_1(M^2_{\tilde{\chi}^-_A}/
m^2_{\tilde{\nu}_X})
\right],
\\
A^{(n)}_L&=&\frac{e}{16 \pi^2}\frac{1}{m^2_{\tilde l_X}}
\left[
 N_{jAX}^{L} N_{iAX}^{L*} J_1(M^2_{\tilde{\chi}^0_A}/m^2_{\tilde{l}_X})
+N_{jAX}^{L} N_{iAX}^{R*} \frac{M_{\tilde{\chi}_A^0}}{m_{l_i}}
J_2(M^2_{\tilde{\chi}^0_A}/m^2_{\tilde{l}_X})
\right],
\\
A^{(c,n)}_R&=&A^{(c,n)}_L|_{L \leftrightarrow R},
\eea
where $M_{\tilde{\chi}^-_A} (M_{\tilde{\chi}^0_A})$ is the chargino (neutralino) mass and $m^2_{\tilde{\nu}_X} (m^2_{\tilde{l}_X})$ is the sneutrino (charged slepton) mass squared.

The functions $I_{1,2}$ and $J_{1,2}$ are defined as follows:
\bea
I_1(r)&=& \frac{1}{12 (1-r)^4} (2+3r-6r^2+r^3+6r \ln r),
\\
I_2(r)&=& \frac{1}{2(1-r)^3} (-3+4r-r^2-2 \ln r),
\\
J_1(r)&=& \frac{1}{12 (1-r)^4} (1-6r+3r^2+2r^3-6r^2\ln r),
\\
J_2(r)&=& \frac{1}{2(1-r)^3} (1-r^2+2r \ln r),
\eea

Finally,
\bea
 C^R_{iAX}& =& -g_2(O_R)_{A1} U^{\nu}_{X,i}, \\
 C^L_{iAX}& = & g_2\frac{m_{l_i}}{\sqrt{2}m_W\cos\beta}(O_L)_{A2}
                    U^{\nu}_{X,i},
\eea
where $A=1,2$ and $X=1,2,3$, and
\bea
N^R_{iAX}&=& -\frac{g_2}{\sqrt{2}} \{
       [-(O_N)_{A2} -(O_N)_{A1} \tan \theta_W] U^l_{X,i}
        + \frac{m_{l_i}}{m_W\cos\beta} (O_N)_{A3} U^l_{X,i+3} \},\\
N^L_{iAX} &=& -\frac{g_2}{\sqrt{2}} \{
           \frac{m_{l_i}}{m_W\cos\beta} (O_N)_{A3} U^{l}_{X,i}
           +2 (O_N)_{A1} \tan \theta_W U^{l} _{X,i+3} \},
\eea
where $A=1,...,4$ and $X=1,...,6$. The different matrices in 
these equations are defined in Appendix B of ref.\cite{Hisano:1996cp}.

\end{document}